\documentclass[twocolumn,amsmath,amsfonts,amssymb,aps,prd,10pt,superscriptaddress,nofootinbib,noeprint,preprintnumbers,floatfix]{revtex4-2}

\usepackage[section]{placeins}
\usepackage{slashed}

\usepackage{tikz}
\usetikzlibrary{decorations.pathreplacing,decorations.markings}
\usepackage{tikz-feynman}

\usepackage{subcaption}
\usepackage{graphicx, color}

\usepackage[letterspace=-10]{microtype} 

\usepackage{bm, amsmath, amsfonts}
\usepackage{multirow, tabularx, dcolumn}
\usepackage{mathtools} 
\usepackage{booktabs}
\usepackage{placeins}
\usepackage{soul} 

\usepackage[utf8]{inputenc} 
\usepackage{hyperref}

\usepackage{xcolor}
\definecolor{jlab_red}{RGB}{192,39,45}
\definecolor{jlab_orange}{RGB}{249,102,0}
\definecolor{jlab_blue}{RGB}{47,122,121}
\definecolor{jlab_green}{RGB}{65,125,10}
\definecolor{jlab_grey}{RGB}{125,125,125}

\hypersetup{%
pdftitle = {eta and eta' meson production in J/psi radiative decays from lattice QCD},
pdfsubject = {QCD},
pdfkeywords = {QCD, Hadron, Charmonium, Physics, Lattice, Meson},
pdfauthor = {Hadron Spectrum Collaboration},
colorlinks = {true},
filecolor = {black},
linkcolor = {jlab_blue},
menucolor = {black},
citecolor = {jlab_green},
urlcolor = {jlab_green},
}{}

\begin{document}

\title{$\eta$ and $\eta'$ meson production in $J/\psi$ radiative decays from lattice QCD}

\author{Mischa~Batelaan}
\email{mbatelaan@wm.edu}
\affiliation{Department of Physics, College of William and Mary, Williamsburg, VA 23187, USA}
\author{Jozef~J.~Dudek}
\email{dudek@jlab.org}
\affiliation{Department of Physics, College of William and Mary, Williamsburg, VA 23187, USA}
\affiliation{\lsstyle Thomas Jefferson National Accelerator Facility, 12000 Jefferson Avenue, Newport News, VA 23606, USA}
\author{Robert~G.~Edwards}
\email{edwards@jlab.org}
\affiliation{\lsstyle Thomas Jefferson National Accelerator Facility, 12000 Jefferson Avenue, Newport News, VA 23606, USA}

\collaboration{for the Hadron Spectrum Collaboration}
\noaffiliation

\date{June 11, 2025}

\begin{abstract}
\noindent
We report on the computation of amplitudes describing the radiative decay processes $J/\psi \to \gamma \, \eta$ and $J/\psi \to \gamma \, \eta'$ from first principles via lattice QCD. Using lattices with two degenerate flavors of light quark and a heavier strange quark where $m_\pi \sim 391$ MeV, we compute three--point correlation functions using optimized meson operators with a range of momenta. The use of optimized operators allows access to the $\eta'$, even though it appears as an excited state, lying above the ground--state $\eta$ in the isoscalar pseudoscalar channel.
Statistically precise signals are obtained in this disconnected process by averaging over large numbers of kinematically equivalent correlators.
We determine transition form--factors as a function of photon virtuality across the timelike region, which describe also the `Dalitz' processes $J/\psi \to e^+ e^- \, \eta^{(\prime)}$. 
Significantly lower magnitudes of transition form--factor for both the $\eta$ and $\eta'$ are found than those extracted from experimental data, and possible explanations for this observation are presented.  
\end{abstract}

\preprint{JLAB-THY-25-4370}

\maketitle

\section{Introduction}\label{intro}

Charmonium radiative decays have become a powerful tool for the study of light hadron spectroscopy, driven by the contemporary dataset of ten billion $J/\psi$ produced directly in $e^+ e^-$ collisions at BESIII, together with significant numbers of $\psi(2S)$, and $h_c, \chi_{cJ}$ produced in $\psi(2S)$ decays.
In particular, the total decay width of $J/\psi$ being small, at the keV level, means that radiative decays to systems of light mesons constitute significant branching fractions that can be measured with good statistics \cite{Jin:2021vct}.

Examples of physics reached from $J/\psi$ radiative decays include observations of isoscalar resonances decaying to $\pi\pi, K\bar{K}$~\cite{BESIII:2015rug, BESIII:2018ubj}, and the recent claim of an  isoscalar \emph{exotic} $J^{PC} = 1^{-+}$ state, the $\eta_1(1855)$, decaying to $\eta \eta'$, which may have an interpretation as a hybrid meson~\cite{BESIII:2022riz}.

Relatively long-lived light mesons can also be produced in isolation, notably the $\eta$ and $\eta'$ in the process ${J/\psi \to \gamma \eta^{(\prime)}}$, where the branching fraction to the $\eta'$ is measured to be around five times larger than that to the $\eta$~\cite{BESIII:2023fai}. The large BESIII dataset has also allowed access to the related `Dalitz' processes in which the emitted photon is \emph{virtual} and \emph{timelike} producing a detected $e^+e^-$ pair, $J/\psi \to e^+ e^- \, \eta^{(\prime)}$~\cite{BESIII:2018qzg, BESIII:2018aao}.  These processes are described in terms of a \emph{transition form--factor} as a function of photon virtuality, $Q^2 = - m_{ee}^2$, with the true radiative decay corresponding to the $Q^2=0$ point.

Radiative decays of the $\psi(2S)$ show a surprisingly large suppression for the $\eta$ with respect to the $\eta'$, with current experimental results having the ratio $\frac{\mathcal{B}(\psi(2S)\to\gamma\eta)}{\mathcal{B}(\psi(2S)\to\gamma\eta')}$ being approximately 30 times smaller than the same ratio for the $J/\psi$ radiative decays \cite{BESIII:2017nnc}. In contrast, recent BESIII results find the corresponding ratio for the $h_c$ decaying to $\gamma \eta$ and $\gamma \eta'$ is much closer to the $J/\psi$ value than the $\psi(2S)$ value~\cite{BESIII:2024xfs}.

\medskip

Radiative decays $J/\psi \to \gamma X$ provide a particularly clean theoretical environment to study light hadron systems, $X$, as compared to hadronic decays $J/\psi \to h X$, owing to the lack of strong rescattering between the photon and $X$. 

The need for the $c\bar{c}$ pair to annihilate to access the light hadron system has led to these often being described as `gluon-rich' production processes, and hence proposed as a good place to look for glueballs. While no clear consensus has arisen for the role of glueball basis states in the spectrum, some of the cleanest modern data on isoscalar meson-meson pairs comes from these processes. Access to different final states (e.g. $\pi\pi, K\bar{K}, \ldots$) through a common production method straightforwardly admits constrained \emph{coupled-channel} analysis, and recent high-statistics BESIII data has been so analyzed in terms of $f_0$ and $f_2$ resonances~\cite{Rodas:2021tyb, Sarantsev:2021ein}.

The $\eta_1(1855)$ state claimed in the BESIII measurement of $J/\psi \to \gamma\,  \eta \eta'$ poses some phenomenological questions – its production in a $J/\psi$ radiative decay signals that it has an $SU(3)$ flavor singlet component, but its decay into $\eta \eta'$ (where the $\eta$ is dominantly octet and the $\eta'$ dominantly singlet) suggests an octet component. This indicates the need for a basis-state mixing pattern different to that of the pseudoscalar $\eta, \eta'$ pair, and furthermore suggests that there should be a \emph{second} isoscalar $1^{-+}$ state, the orthogonal admixture of octet and singlet, which should also appear in this radiative decay. Experimentally there is no sign of a second resonance. 

In principle the simplest radiative decays are those with just a single long-lived meson in the final state, such as $J/\psi \to \gamma \, \eta$ or $J/\psi \to \gamma \, \eta'$. Strong arguments can be made that in these processes the photon originates from the initial-state charm quarks rather than being emitted by any of the light quarks in the final state. The extremely narrow width of the $J/\psi$ suggests that annihilation into only gluons, leading to light hadrons, is heavily suppressed. On the other hand, emission of a photon from the $J/\psi$ could place the charmonium system into a virtual pseudoscalar state, and from the example of the $\eta_c$ having a total width in the tens of MeV, we know that annihilation of such a system is relatively unsuppressed. Further support for this argument comes from the measured rate for $J/\psi \to \gamma  \pi^0$ which is around 150 times smaller than the rate for $J/\psi \to \gamma  \eta'$. In the limit of good isospin symmetry in QCD, the $\gamma \pi^0$ decay requires the photon to couple to a light quark in the final-state system in order to change the isospin from $I=0$ to $I=1$.

\medskip

Attempts to describe $J/\psi \to \gamma \, \eta$ and $J/\psi \to \gamma \, \eta'$ theoretically within QCD have a long history. In Ref.~\cite{Novikov:1979uy}, the authors argue that a sum-rule for the process $\gamma^\star \to \gamma \eta^{(\prime)}$ proceeding through a charm-quark loop is largely saturated by the contribution of the $J/\psi$ to $\gamma^\star$, while the dynamics associated with the $\eta^{(\prime)}$ production is entirely described by a single gluonic matrix element that can be set by PCAC (for the $\eta$) or by other low-energy sum-rules (for the $\eta'$). Other approaches treat the intermediate gluonic system perturbatively (e.g.~\cite{Korner:1982vg}), or in terms of multipoles~\cite{Kuang:1990kd}, but in all cases the non-perturbative aspects are parameterized, and if they are determined at all, it is in terms of some other measured quantity, rather than from first-principles in QCD.

\medskip

Lattice QCD offers the opportunity to compute radiative charmonium decays from first-principles, but these calculations are particularly challenging, even in the simplest case of a single QCD-stable light meson in the final state, $J/\psi \to \gamma\,  \eta^{(\prime)}$. The challenges come in several forms:
The contributing Wick diagrams are \emph{disconnected}, which typically leads to noisier signals. It is required to have two light flavors of quark and a third heavier strange quark in order to have a true representation of the $\eta$, $\eta'$ system. The $\eta'$ appears as an \emph{excited state} in the isoscalar pseudoscalar channel and hence, unlike the \mbox{ground state $\eta$}, cannot simply be accessed by going to large time separations in correlation functions. The large mass difference between the $J/\psi$ and the light meson in the final state means that the region of photon virtuality near zero can only be accessed by considering mesons having large values of three-momentum, which typically leads to degraded signal--to--noise.
 
These challenges indicate that showing that sufficiently good signals can be obtained in a lattice QCD calculation of $J/\psi \to \gamma\,  \eta^{(\prime)}$ is a precursor to computing processes like $J/\psi \to \gamma\,  \pi \pi$, $J/\psi \to \gamma \, K \bar{K} $ and others in which $f_J$ resonances can appear. These more complex processes require the additional application of finite-volume formalism as described in Refs.~\cite{Briceno:2021xlc}(and references therein) and applied to other processes in Refs.~\cite{Briceno:2015dca, Briceno:2016kkp, Radhakrishnan:2022ubg, Ortega-Gama:2024rqx, Alexandrou:2018jbt, Leskovec:2025gsw}.

Previous lattice calculations intending to investigate $J/\psi \to \gamma \, \eta^{(\prime)}$ used a maximum of two quark flavors and considered the ground state pseudoscalar only~\cite{Jiang:2022gnd, Shi:2024fyv}. As such they lacked separate access to the $\eta$ and $\eta'$ and instead used phenomenological arguments to determine the rates to the two different final states. In addition, these exploratory calculations sampled only a limited set of kinematic configurations, having the $J/\psi$ at rest and the light pseudoscalar with several values of three-momentum.

\bigskip

In this paper we will report on a first calculation of \emph{both} $J/\psi \to \gamma \, \eta$ and $J/\psi \to \gamma \, \eta^\prime$ in lattice QCD, using ${2+1}$ flavor anisotropic lattices with $m_\pi \sim$ 391 MeV on which the light meson and charmonium spectrum have previously been studied extensively. We will make use of the full power of distillation~\cite{HadronSpectrum:2009krc} in computing three-point functions using \emph{optimized meson operators}. These operators, obtained by variational analysis of a matrix of two-point functions, cause rapid relaxation in Euclidean time to desired states, and allow for access to the $\eta'$ as an excited state.
We will present results extracted from three-point functions in a very large number of kinematic configurations with meson momenta up to $|\mathbf{p}|^2 = 6 \left(\tfrac{2\pi}{L}\right)^2$, having opted to appropriately average over correlators related by lattice symmetries to improve statistics. We will show statistically precise signals which lead to transition form--factors computed at seventy $Q^2$ values from deeply timelike through to slightly spacelike. The $Q^2$ dependences will be described by suitable analytic forms and the $J/\psi$ decay rates from the $Q^2=0$ point will be presented. 

\vspace{5mm}

The structure of the remainder of the manuscript is as follows: After defining transition form--factors (Sec.~\ref{raddecays}), we will describe the computational details for this calculation (Sec.~\ref{numdetails}), before presenting methods applied to access mesons in two-point correlation functions (Sec~\ref{twoptfunctions}). Sec.~\ref{threeptfunc} describes the extraction of transition matrix elements from three-point correlation functions. Sec.~\ref{formfactors} presents the results of our calculation, including the $Q^2$ dependence of the form--factors, and the radiative decay rates. Discussion and outlook are presented in Sec.~\ref{conclusion}, before a series of Appendices fill in certain details.


\section{$J/\psi$ Radiative Decay to stable Pseudoscalar mesons}\label{raddecays}

At lowest order in the electromagnetic coupling, $\alpha$, the radiative decays $J/\psi\to\gamma \, \eta^{(\prime)}$, and the timelike Dalitz decays $J/\psi\to e^+ e^- \, \eta^{(\prime)}$, can be described by  matrix elements,
\begin{equation*}
\label{eq:1}
  \big\langle \eta^{(\prime)}(\mathbf{p}') \big| j_{\mathrm{em}}^{\mu}(0) \big| J/\psi(\mathbf{p}, \lambda) \big\rangle \, ,
\end{equation*}
which have a decomposition in terms of a single Lorentz-invariant form--factor,
\begin{align}
  \label{eq:K}
  \begin{split}
  &\big\langle \eta^{(\prime)}(\mathbf{p}') \big| j_{\mathrm{em}}^{\mu}(0) \big| J/\psi(\mathbf{p}, \lambda) \big\rangle =  \\
  &\qquad \qquad \qquad \epsilon^{\mu\nu\rho\sigma}\, p'_{\nu}\, p_{\rho}\, \epsilon_{\sigma}(\mathbf{p}, \lambda) \, F_{\psi \eta^{(\prime)}}(Q^2) \, .
  \end{split}
\end{align}
The dependence upon the $J/\psi$ helicity is embedded in the polarization vector, while the form--factor depends only on the virtuality of the photon, 
\begin{equation*}
 Q^2 = | \mathbf{p}' - \mathbf{p} |^2 - (E_{\eta^{(\prime)}} - E_\psi)^2 \, ,
\end{equation*}
with the real-photon decay process corresponding to $F_{\psi \eta^{(\prime)}}(0)$.

As argued in the introduction, the dominant contribution to the matrix element comes from the electromagnetic current coupling to a charm quark in the $J/\psi$, so that we can approximate $j^\mu_{\mathrm{em}} =   \bar{c}\gamma^\mu c$ in units of the charm--quark charge, $\tfrac{2}{3} e$. The corresponding partial decay rates can then be expressed as
\begin{equation}
\label{eq:width}
\Gamma \big(J/\psi \to \gamma \, \eta^{(\prime)} \big) = \frac{4}{27} \alpha \, |\mathbf{q}|^3 \, \big|F_{\psi \eta^{(\prime)}}(0)\big|^2,
\end{equation}
where $|\mathbf{q}|$ is the magnitude of the photon momentum in the rest frame of the $J/\psi$.

Additional support for the argument that coupling to the charm quark in the $J/\psi$ dominates over coupling to light quarks comes from examination of the timelike form--factor\footnote{Defined in a manner which differs from Eqn.~\ref{eq:K} only by a normalization~\cite{Landsberg:1985gaz}.} extracted from $J/\psi\to e^+ e^- \, \eta$ data in Ref.~\cite{BESIII:2018qzg} (shown in the erratum's Figure 2).
In general, assuming isospin symmetry in QCD, in the timelike region this process would have a discontinuity in $Q^2$ at the opening of the $\pi\pi\pi$ channel, and might be expected to show enhancements at the location of resonances appearing in that channel, such as the $\omega(782)$. The data points show no significant departures from a smooth behavior for $Q^2 \gtrsim -5.7 \,\mathrm{GeV^2}$, except for a single--bin fluctuation near the $\omega$ mass\footnote{Which the authors of Ref.~\cite{BESIII:2018qzg} associate with the $\rho$ resonance, even though that would require isospin violation.}. %


\section{Calculational Details}\label{numdetails}

For this first calculation of $J/\psi$ radiative decays to $\eta$ \emph{and} $\eta'$ separately, we will make use of a single ${2+1}$ flavor ensemble of anisotropic clover lattices. This ensemble is part of a set that have been used extensively to study the light hadron spectrum, as well as the charmed meson spectrum and charmonium.
Parameters in the gauge and fermion actions were tuned to generate an anisotropy ${\xi = {a_s}/{a_t} \approx 3.5}$ with a spatial lattice spacing, $a_s \approx 0.12 \,\textrm{fm}$~\cite{HadronSpectrum:2008xlg}.
The lattices feature two degenerate light-quarks which are heavier than physical, and a strange quark which is approximately at the physical strange mass -- the pion mass is found to be $m_{\pi} \approx 391\,\textrm{MeV}$. We use a single volume in this first calculation, $(L/a_{s})^3\times (T/a_{t}) = 20^3\times 256$, where $m_{\pi}L \approx 5$. 
All quantities with physical units in this paper will be scale set using the ratio of the physical mass of the $\Omega$ baryon to its mass determined on this lattice, yielding an inverse temporal lattice spacing, $a_t^{-1} = \frac{m_{\Omega}^{\textrm{phys}}}{a_t m_{\Omega}^{\textrm{lat.}}} = 5.666 \,\textrm{GeV}$ \cite{Edwards:2012fx}.

On the current lattice the $\eta$ mass is found to be $m_\eta = 0.1030(4)\, a_t^{-1} = 584(2)\, \mathrm{MeV}$ and the $\eta'$ mass $m_{\eta'} = 0.167(2)\, a_t^{-1} = 945(9)\, \mathrm{MeV}$. Considering the larger than physical light--quark mass it is perhaps slightly surprising that the $\eta'$ mass is found to be so close to the physical $\eta'$ mass -- we will return to this observation later in this paper when interpreting the results of the current calculation\footnote{In Ref.~\cite{Dudek:2013yja}, a lattice almost the same as this one was used, just with a shorter time-length, $T/a_t = 128$, but no significant differences with respect to the current lattice are observed in the light meson sector. }.

Non-dynamical charm quarks are introduced using the same relativistic clover action as the light quarks, with the quark mass tuned to reproduce the physical $\eta_c$ mass. Ref.~\cite{HadronSpectrum:2012gic} presents the charmonium spectrum on lattices with the same action, but on two different volumes, computed neglecting the possibility of $c\bar{c}$ annihilation\footnote{A tiny effect for the almost--stable $J/\psi$.}. Including also the current lattice volume, the charmonium spectrum appears in Ref.~\cite{Wilson:2023anv}, where the $J/\psi$ mass was found to be $m_\psi = 0.53715(5)\, a_t^{-1} = 3043.5(3)\, \mathrm{MeV}$.

As in Refs.~\cite{Dudek:2013yja, Wilson:2023anv}, \emph{distillation}~\cite{HadronSpectrum:2009krc} smearing with $N_{\text{vec}}=128$ eigenvectors is used in the construction of operators interpolating hadron states.

Matrices of two--point correlation functions are computed using a basis of fermion--bilinear operators with the quantum numbers of $\eta^{(\prime)}$ and $J/\psi$ at three--momentum values up to $|\mathbf{p}|^2 = 6 \left(\tfrac{2\pi}{L} \right)^2$. Variational analyses of these yield energies for the $\eta$, $\eta'$ and $J/\psi$, allowing us to establish that these mesons on the current lattice satisfy consistent relativistic dispersion relations. The variational analysis also supplies the particular linear combination of basis operators (at each momentum) that optimally interpolates each state -- these will be used in the construction of three--point correlation functions.

\begin{figure}
  \centering
  \begin{tikzpicture}[
    thick,
    decoration={
      markings,
      mark=at position 0.5 with {\arrow{latex}}
    }
    ]

    \draw[postaction={decorate}] (-2,1) to[out=0,in=90] (-1,0);
    \draw[postaction={decorate}] (-1,0) to[out=-90,in=0] (-2,-1);
    
    \draw[decorate,decoration={snake}] (-1,0) -- (0,1);
    
    \draw[postaction={decorate}] (2,-1) to[out=180,in=-90] (1,0);
    \draw[postaction={decorate}] (1,0) to[out=90,in=-180] (2,1);
    
    \node at (-2.3,1) {$c$};
    \node at (-2.3,-1) {$\bar{c}$};
    \node at (0.3,1.2) {$\gamma$};
    \node at (2.3,1) {$q$};
    \node at (2.3,-1) {$\bar{q}$};
    
    \node at (-2.5,0) {$J/\psi$};
    \node at (2.5,0) {$\eta^{(\prime)}$};
  \end{tikzpicture}
  \caption[]{\label{fig:tikz-psice-gamma-eta} Wick diagram structure required to compute three--point correlation functions. Left--half of diagram corresponds to a charm--quark generalized perambulator, while the right--half is either a light--quark or strange--quark perambulator.}
\end{figure}
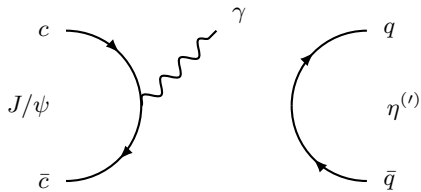

The required three--point correlation functions feature an unsmeared vector current, which appears as an insertion on a charm--quark propagation object, something we refer to as a \emph{generalized perambulator}. The required Wick contraction is illustrated in Figure~\ref{fig:tikz-psice-gamma-eta}, where the rightmost perambulator will in some diagrams be a light quark and in others a strange quark, reflecting the presence in the $\eta$ and $\eta'$ of both hidden--light and hidden--strange components.
The charm--quark generalized perambulators are computed using inversions off time--sources placed every 32 timeslices across the lattice together with two additional sources. At the current insertion, all 16 gamma--matrices\footnote{The clover-improved quark action \cite{HadronSpectrum:2008xlg,Edwards:2008ja} used in this work requires a tree--level $\mathcal{O}(a)$-improvement term to be included in the vector current, featuring gamma--matrix structures other than just $\gamma^\mu$, which we will address in Section~\ref{threeptfunc}.} were computed with momenta up to $\mathbf{q} = \left(\tfrac{2\pi}{L}\right) [0,0,4]$, but with \emph{only a single momentum direction} computed for each magnitude of momentum\footnote{More precisely each `type' of momentum, e.g. $\left(\tfrac{2\pi}{L}\right) [1,2,2]$ and $\left(\tfrac{2\pi}{L}\right) [0,0,3]$ are both computed.}.
The light and strange perambulators had previously been computed on every timeslice of the lattice for use in prior calculations. All propagation objects were computed on 288 configurations, and three--point correlation functions are averaged over the 10 charm sources.

\section{Two--point Functions}\label{twoptfunctions}

As mentioned earlier, the kinematics of the $J/\psi \to \gamma \, \eta^{(\prime)}$ process in typical lattice volumes are such that to access the region where $Q^2 \approx 0$, we require mesons to have significantly nonzero three--momentum. As such we need to consider first two--point correlation functions featuring the $J/\psi$ and separately the light isoscalar pseudoscalars at several values of momentum.

The $J/\psi$ has $J^{PC}$ quantum numbers of $1^{--}$, and appears at rest in the $T_1^-$ irrep of cubic symmetry along with $3^{--}$ and higher spins. At non-zero momenta, the symmetry group is reduced, and some of the relevant irreducible representations also contain $1^{+-}$ quantum numbers, corresponding to the $h_c$ meson.

Operators of $\sum_\mathbf{x} e^{i \mathbf{p} \cdot \mathbf{x} } \big(\bar{\psi} \mathbf{\Gamma} \psi\big)(\mathbf{x})$ structure, transforming irreducibly under the reduced cubic symmetry group for three--momentum $\mathbf{p} = \frac{2\pi}{L} \mathbf{n}$ are constructed following the approach presented in detail in Ref.~\cite{Thomas:2011rh}. In short, this method projects operators which at rest would have definite $J^P$ into definite \emph{helicity}, $\lambda$, at momentum $\mathbf{p}$,
\begin{equation*}
\label{eq:6}
\left[ \mathcal{O}^{[J^P]}_{\lambda}(\mathbf{p}) \right]^{\dagger} = \sum_m D^{(J)}_{m \lambda}(R_\mathbf{p}) \left[ \mathcal{O}^{[J^P]}_m(\mathbf{p}) \right]^{\dagger}\, ,
\end{equation*}
and then \emph{subduces} into irreducible representations, $\Lambda$, of the reduced symmetry of the boosted lattice~\cite{Moore:2005dw},
\begin{equation*}
\label{eq:7}
\left[ \mathcal{O}^{[J^P, |\lambda|]}_{\Lambda,\mu}(\mathbf{p}) \right]^{\dagger} = \sum_{\hat{\lambda}=\pm|\lambda|} S^{\tilde{\eta}, \hat{\lambda}}_{\Lambda, \mu} \left[ \mathcal{O}^{[J^P]}_{\hat{\lambda}}(\mathbf{p}) \right]^{\dagger}.
\end{equation*}
The subduction coefficients are presented in Ref.~\cite{Thomas:2011rh} where the set of $J^P$ appearing in each irrep $\mathbf{n} \Lambda$ can also be found\footnote{$\tilde{\eta} = P (-1)^J$ is a good quantum number in the $\lambda=0$ case.}. We summarize the content of the irreps featuring $J^{PC} = 1^{--}$ that we will use in Table~\ref{tab:subduction}.

\begin{table}

\begin{tabular}{c|cll}
$\mathbf{n}\, \Lambda$ 		& dim.		& helicity content		& $J^{PC}$ content \\
\hline
$[000]\, T_1^-$				& 3						& --						& $\bm{1^{--}}, 3^{--} \ldots$ \\[1ex]
$[100]\, A_1$ 				& 1						& $0, \pm4 \ldots$			& $\mathit{0^{+-}}, \bm{1^{--}}, \mathit{2^{+-}}, 3^{--} \ldots$ \\
$[100]\, E_2$ 				& 2						& $\pm1, \pm3 \ldots$		& $\bm{1^{--}}, 1^{+-}, 2^{--},\mathit{2^{+-}}, 3^{--} \ldots$ \\[1ex]
$[110]\, A_1$ 				& 1						& $0, \pm2, \pm4 \ldots$	& $\mathit{0^{+-}}, \bm{1^{--}}, 2^{--}, \mathit{2^{+-}}, 3^{--} \ldots$ \\
$[110]\, B_1$ 				& 1						& $\pm1, \pm3 \ldots$		& $\bm{1^{--}}, 1^{+-}, 2^{--}, \mathit{2^{+-}}, 3^{--} \ldots$ \\
$[110]\, B_2$ 				& 1						& $\pm1, \pm3 \ldots$		& $\bm{1^{--}}, 1^{+-}, 2^{--}, \mathit{2^{+-}}, 3^{--} \ldots$ \\[1ex]
$[111]\, A_1$ 				& 1						& $0, \pm3 \ldots$			& $\mathit{0^{+-}}, \bm{1^{--}}, \mathit{2^{+-}}, 3^{--} \ldots$ \\
$[111]\, E_2$ 				& 2						& $\pm1, \pm2, \pm4 \ldots$	& $\bm{1^{--}}, 1^{+-}, 2^{--},\mathit{2^{+-}}, 3^{--} \ldots$ \\[1ex]
$[200]\, A_1$ 				& 1						& $0, \pm4 \ldots$			& $\mathit{0^{+-}}, \bm{1^{--}}, \mathit{2^{+-}}, 3^{--} \ldots$ \\
$[200]\, E_2$ 				& 2						& $\pm1, \pm3 \ldots$		& $\bm{1^{--}}, 1^{+-}, 2^{--},\mathit{2^{+-}}, 3^{--} \ldots$ \\[1ex]
\end{tabular}
\caption{\label{tab:subduction}Irreps containing $1^{--}$ $J/\psi$ to be used in three--point functions. The dimension of the irrep is reflected in the number of `rows', $\mu = 1 \ldots \mathrm{dim}$. $A_1$ irreps contain the $\lambda=0$ component of the vector, while the two rows of $E_2$, and the $B_1, B_2$ pair, contain the sum and difference of the $\lambda=\pm1$ components. Other $J^{PC}$ appearing in these irreps are indicated in the final column, where entries in italics are \emph{exotic} and do not appear as low-lying resonances.}
\end{table}

\smallskip

For the $J/\psi$ case, our basis of operators uses only charm--quark fermion bilinears, with independent operators built from up to three gauge--covariant derivatives at rest and up to two for nonzero momentum~\cite{Dudek:2010wm}. The two--point correlation functions computed in this case are built assuming only connected diagrams, disallowing charm--quark annihilation~\cite{HadronSpectrum:2012gic, Wilson:2023anv}.

\smallskip

In the $\eta$, $\eta'$ sector, the same operator construction approach is used (in different irreps to those above, reflecting the $J^{PC} = 0^{-+}$ quantum numbers), but the flavor basis is broader, 
\begin{equation*}
\mathcal{O}^\ell = \tfrac{1}{\sqrt{2}}(\bar{u} \mathbf{\Gamma} u + \bar{d}\mathbf{\Gamma} d),\quad \mathcal{O}^s = \bar{s} \mathbf{\Gamma} s\, ,
\end{equation*}
and annihilation diagrams \emph{are} included, allowing hidden--light, hidden--strange mixing~\cite{Dudek:2011tt,Dudek:2013yja}. In this case the relevant irreps are $A_1^-$ at rest and $\mathbf{n}A_2$ for all nonzero momenta.

\medskip

Independently in each relevant irrep, we construct a two--point correlator matrix built from the operator basis $\mathcal{O}_i$, $i=1 \ldots N$,
\begin{equation*}
C_{ij}(t) = \big\langle 0 \big| \mathcal{O}^{}_i(t)\,  \mathcal{O}^{\dagger}_j(0) \big| 0 \big\rangle \, ,
\end{equation*}
and to determine the energy spectrum we use a variational method, based upon solving a generalized eigenvalue problem, 
\begin{equation}
\label{eq:GEVP}
\mathbf{C}(t) \, v_\mathfrak{n} = \lambda_\mathfrak{n}(t, t_0) \, \mathbf{C}(t_0) \, v_\mathfrak{n} \, ,
\end{equation}
with an implementation as presented in Ref.~\cite{Dudek:2007wv}. This diagonalises the correlator matrix and returns a set of eigenvalues, $\lambda_\mathfrak{n}$, from which the discrete energy spectrum $E_\mathfrak{n}$ can be extracted, and eigenvectors, $v_\mathfrak{n}$, which give the linear combination of operators which couples best to each state, $|\mathfrak{n}\rangle$. An \emph{optimized operator} for state $|\mathfrak{n}\rangle$ is then constructed as\footnote{In fact a small adjustment in normalization is made to ensure that $\langle \mathfrak{n} | \Omega^\dag_{\mathfrak{n}}(0) |0 \rangle = \sqrt{ 2 E_\mathfrak{n}}$ for states having normalization $\langle \mathfrak{n} | \mathfrak{n} \rangle = 2 E_\mathfrak{n}$. This is required for finite values of $t_0$ because $\lambda_\mathfrak{n}(t, t_0)$ tends to a finite constant smaller than unity for $t \gg t_0$.}
\begin{equation*}
\label{eq:9}
\Omega_\mathfrak{n}^{\dagger} =  e^{-E_\mathfrak{n} t_{0} / 2} \sum\nolimits_{i}(v_\mathfrak{n})_{i} \mathcal{O}^{\dagger}_i \, .
\end{equation*}
These optimized operators can be used in three-point correlation functions featuring the vector current in order to efficiently relax to the desired states, even if they are excited states, at small Euclidean times, as first demonstrated for the case of distillation--based operators in Ref.~\cite{Shultz:2015pfa}.



\medskip

\begin{figure}
\centerline{\includegraphics[width=.9\columnwidth]{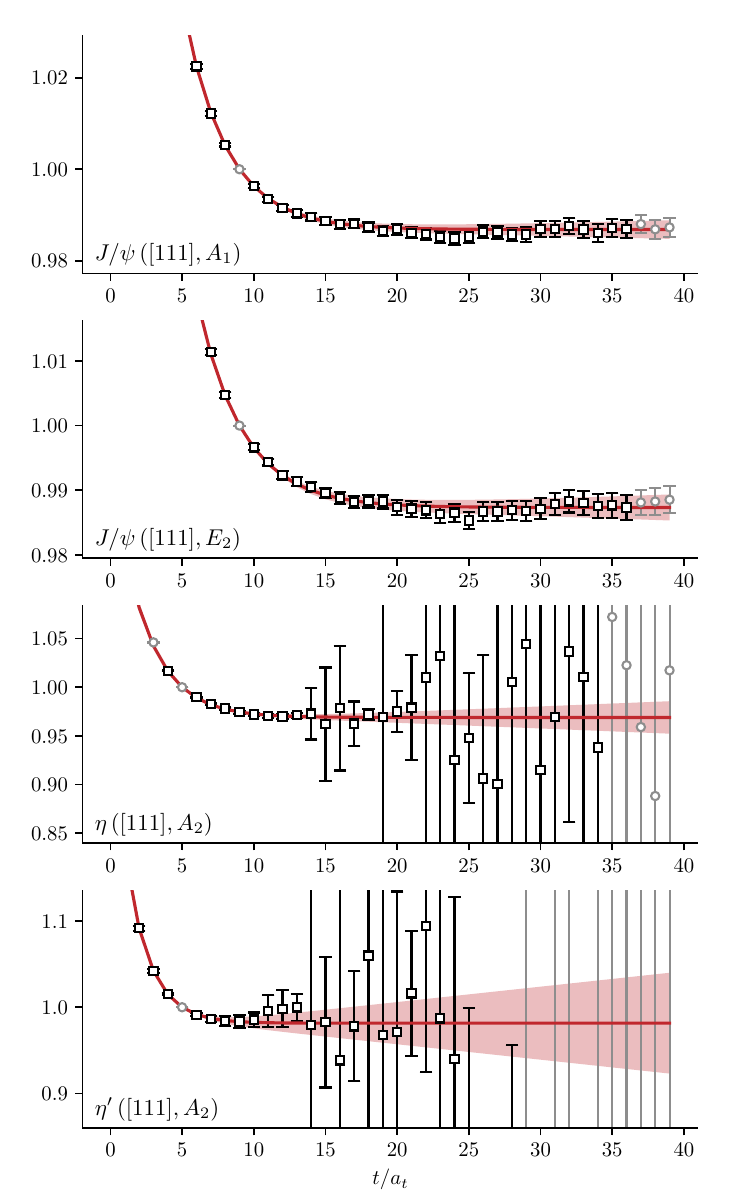}} 
\caption[]{\label{fig:prin-corr-111} 
Eigenvalues of Eqn.~\ref{eq:GEVP} weighted by the leading time-dependence, $e^{E_{\mathfrak{n}}(t-t_{0})}\lambda_{\mathfrak{n}}(t,t_0)$, for $\mathbf{n}=[111]$. Ground--state in two $c\bar{c}$ irreps (upper two panels), and ground and first--excited state in an isoscalar irrep (lower two panels). Colored bands show the result of a two--exponential fit for a single time--window (grey points outside the time-window). 
}
\end{figure}

As an example of the procedure described above, in Figure~\ref{fig:prin-corr-111} we show results from the case $\mathbf{n}=[111]$. The upper two panels show the eigenvalues corresponding to the ground-state $J/\psi$ in the $c\bar{c}$ $A_1$ and $E_2$ irreps, while the lower two panels show two eigenvalues in the isoscalar $A_2$ irrep, corresponding to the $\eta$ (ground state) and $\eta'$ (first excited state). In practice we make use of energies constructed from a weighted average over several time--windows via the Akaike Information Criterion (AIC) as presented in Ref.~\cite{Jay:2020jkz}.

\begin{figure}
\centerline{\includegraphics[width=.95\columnwidth]{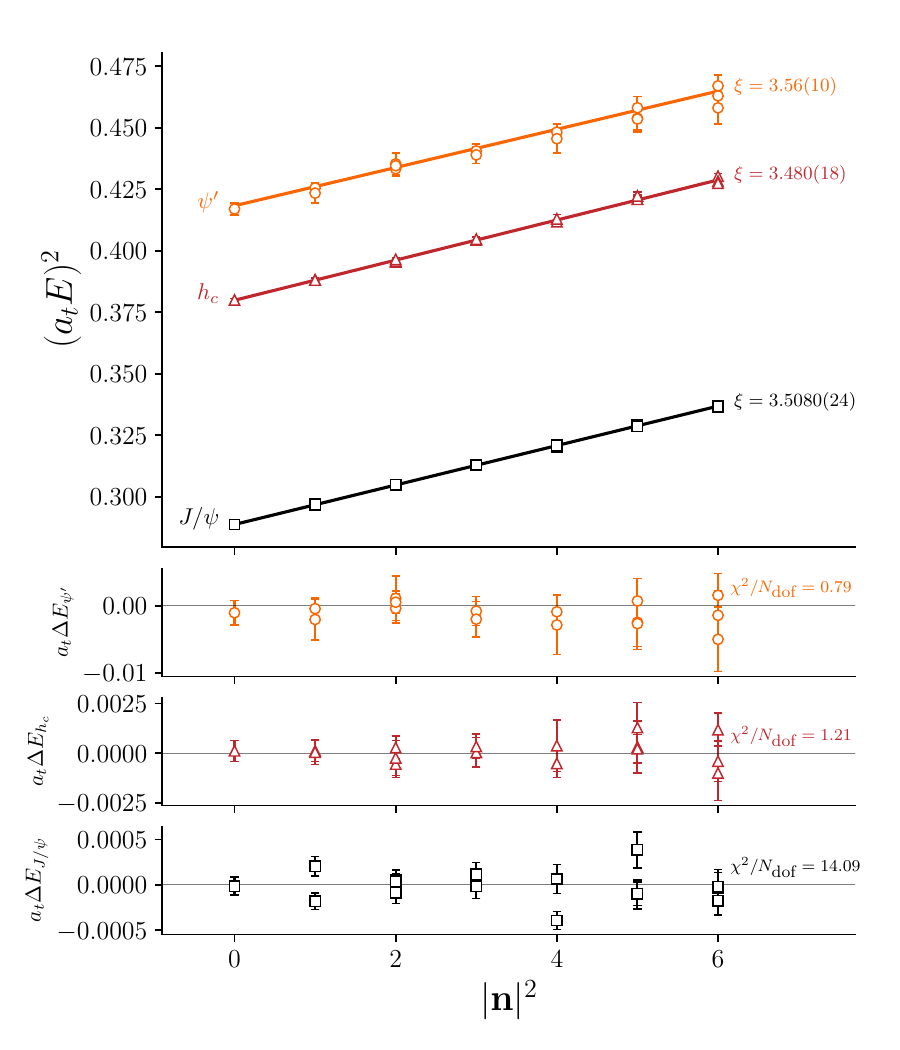}} 
\caption[]{\label{fig:disp-psi} Energies of $J/\psi$, $h_c$, and $\psi'$ mesons for momenta up to $|\mathbf{n}|^2=6$. Where a momentum has more than one energy value, these reflect the energies determined in different irreps (e.g. $A_1, B_1, B_2$ at $\mathbf{n}=[111]$). Lines show dispersion relation fits according to Eqn.~\ref{eq:dispersion}, with departures from these fits shown below.}
\end{figure}

For $c\bar{c}$ irreps at all momenta considered, the variational approach yields good signals for several states, which depending upon the irrep can correspond to the  $J/\psi$, $h_c$ and $\psi'$. The extracted energies of these states as a function of momentum are presented in Figure~\ref{fig:disp-psi}, along with a fit to the continuum--like relativistic dispersion relation,
\begin{equation}
\label{eq:dispersion}
(a_t\, E)^2 = (a_t \, m)^2 + \frac{1}{\xi^2} \left( \frac{2\pi}{L/a_s}\right)^2 |\mathbf{n}|^2 \, ,
\end{equation}
where the hadron mass and the anisotropy, $\xi$, are allowed to vary. We observe that the determined anisotropies for each charmonium hadron are broadly in agreement.

\begin{figure}
\centerline{\includegraphics[width=.95\columnwidth]{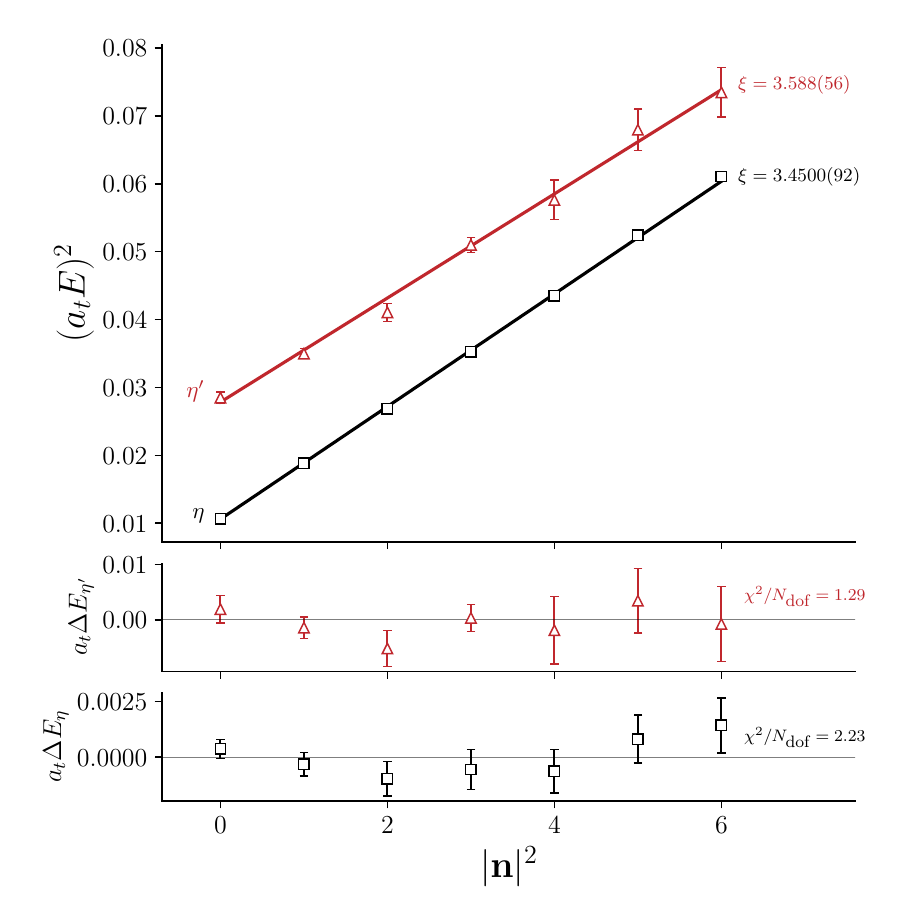}} 
\caption[]{\label{fig:disp-eta} Energies of $\eta$, $\eta'$ mesons for momenta up to $|\mathbf{n}|^2=6$. Lines show dispersion relation fits according to Eqn.~\ref{eq:dispersion}, with departures from these fits shown below.}
\end{figure}

Figure~\ref{fig:disp-eta} similarly shows the dispersion relations for the $\eta$ and $\eta'$, where we see anisotropy values in agreement with those determined in charmonium, as given in Figure~\ref{fig:disp-psi}. For the rest of the analysis we will take $\xi=3.51$, and consider variation of this value in the range 3.45--3.59 as a source of systematic error. Extension of this analysis to higher values of charmonium momentum (not used in the three--point function work to follow) is presented in Appendix~\ref{decay-const} together with consideration of the $J/\psi$ decay constant.

\section{Three--point Functions}\label{threeptfunc}

In order to access the matrix elements needed to describe $J/\psi \to \gamma\,  \eta^{(\prime)}$ we compute three--point correlation functions of the form,
\begin{equation*}
\label{eq:13}
  C(t, \Delta t) =  \big\langle 0 \big|  \Omega_{\eta^{(\prime)}}(\Delta t) \; j(t) \; \Omega^{\dagger}_{\psi}(0) \big| 0 \big\rangle \, ,
\end{equation*}
and owing to the use of optimized operators (as defined in the previous section) to interpolate the $J/\psi$ and the $\eta^{(\prime)}$, even at modest time-separations we expect dominance by the desired states,
\begin{equation*}
  \label{eq:14}
  C(t, \Delta t) = e^{-E_{\eta^{(\prime)}} (\Delta t - t)} e^{- E_\psi t} \frac{\langle \eta^{(\prime)} | j(0) | \psi \rangle}{\sqrt{2 E_{\eta^{(\prime)}} \, 2 E_\psi}} + \ldots \, ,
\end{equation*}
with excited state contributions in the ellipsis. The following simple rescaling yields the desired matrix element as a time--independent constant up to exponentially decaying corrections at source and/or sink coming from (suppressed) excited states,
\begin{align*}
  \tilde{C}(t, \Delta t) &\equiv \frac{\sqrt{2 E_{\eta^{(\prime)}} \, 2 E_\psi} }{ e^{-E_{\eta^{(\prime)}} (\Delta t - t)} e^{- E_\psi t} } \,  C(t, \Delta t) \\
  &= \langle \eta^{(\prime)} | j(0) | \psi \rangle + \ldots \, .
\end{align*}
The time-dependence of objects having the structure of $\tilde{C}(t, \Delta t)$, for a fixed choice of $\Delta t$, can be fitted using the functional form,
\begin{equation*}
\label{eq:threePtFit}
\tilde{C}(t, \Delta t) = J + A_{\text{src}}\,  e^{-\delta E_{\text{src}} t}   + A_{\text{snk}} \, e^{-\delta E_{\text{snk}} ( \Delta t - t )} \, ,
\end{equation*}
where $J$ is the desired matrix element value, and $A_{\text{src}},\, A_{\text{snk}},\, \delta E_{\text{src}},\, \delta E_{\text{snk}}$ are additional free parameters describing excited state contributions\footnote{We are tacitly assuming that the use of optimized operators at source and sink has reduced the contribution of excited-state--to--excited-state transitions to a negligible value.}.
Our approach is to consider many fits over a range of time-windows, allowing an exponential at the source, at the sink, at both, or at neither, and from all these fits, we combine up to 30 having the largest values of AIC using the weighted averaging procedure presented in Ref.~\cite{Jay:2020jkz}, in order to obtain a conservatively determined value of $J$.

\begin{figure*}[htbp]
  \centerline{\includegraphics[width=\textwidth]
    {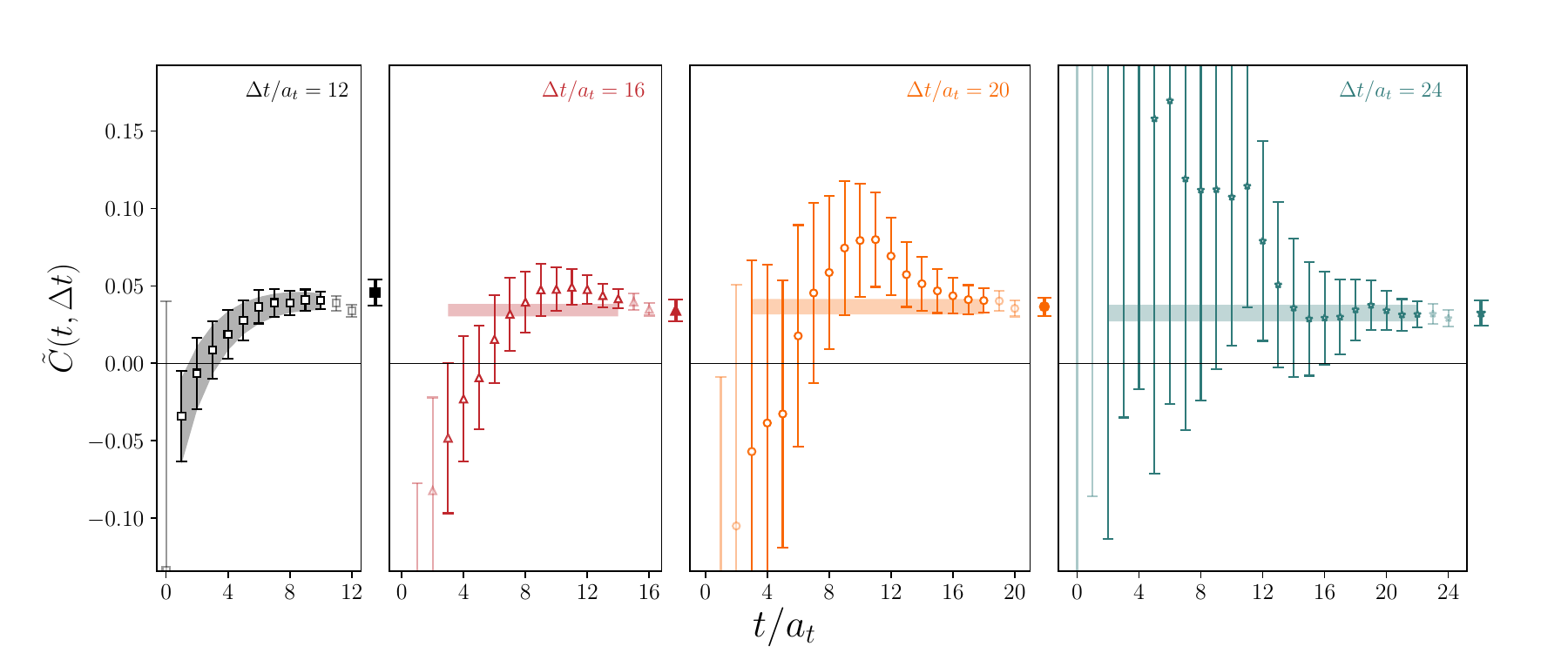}
  }
\caption[]{\label{fig:proj0-dt-comp1} An example $\tilde{C}(t, \Delta t)$ for one particular correlation function constructed to describe $J/\psi \to\gamma\, \eta'$. The sink and source operators have momenta $\mathbf{n}_{\eta'} = [0,0,\text{-}2]$, $\mathbf{n}_{\psi} = [0,0,2]$ respectively, so that the virtuality of the photon is $Q^2 = 0.69 \, \mathrm{GeV}^2$. For each $\Delta t$ value, the band shows the highest AIC--value single time-window fit, and the bold data--point shows the AIC--weighted average value of $J$.}
\end{figure*}

Increasing source--sink separation $\Delta t$ suppresses excited-state contributions at the cost of increased statistical uncertainty. Figure~\ref{fig:proj0-dt-comp1} illustrates the dependence on $\Delta t$ of our correlation functions, showing one particular case for $\Delta t/a_t = 12, 16, 20$ and $24$, along with the highest AIC--value single time--window fit, and the AIC--weighted average value of $J$. The fact that the four extracted values are statistically compatible helps to justify the assumption made in the last footnote.

\medskip

As anticipated in the introduction, correlation functions in this sector are relatively noisy, and to obtain usable signals it is advantageous to average over multiple equivalent determinations. One way to achieve this is to consider correlation functions which are related to each other by rotations which leave the (boosted) lattice invariant\footnote{The data presented in Figure~\ref{fig:proj0-dt-comp1} are the result of this kind of averaging over two related correlation functions.}. To make this practical, one should decompose the vector current in terms of the same irreducible representations as the meson operators. Once this is done, the basic correlation function to be computed carries the following labels:
\begin{equation}
\label{eq:3ptIrrepLabels}
  \big\langle 0 \big|  \, 
  \Omega_{\eta^{(\prime)}}^{\mathbf{p}' \Lambda'\mu'}(\Delta t) 
  \,\, j^{\mathbf{q} \Lambda_{\gamma} \mu_{\gamma}}(t) 
  \, \left[\Omega_\psi^{\mathbf{p}\Lambda \mu}\right]^\dag\!\!(0) \,  
  \big|0 \big\rangle.
\end{equation}
The three--point functions we need to calculate contain a charm--quark annihilation, so we treat the current insertion operator as an annihilation operator, requiring a hermitian conjugate with respect to the subduced helicity operators previously presented for meson creation,
\begin{align*}
  j^{\mathbf{q}\Lambda_{\gamma} \mu_{\gamma}} &=\!\!\! \sum_{\hat{\lambda} = \pm |\lambda|}\!\!\! \left[ S^{\tilde{\eta}, \hat{\lambda}}_{\Lambda_{\gamma}, \mu_{\gamma}}\right]^{*}  \sum_m\! D^{(1)\,*}_{m, \hat{\lambda}}(R_\mathbf{q}) \; (-i) \boldsymbol{\epsilon}^{*}(\mathbf{0},m) \cdot \mathbf{j}(\mathbf{q}) \, ,
\end{align*}
where here we've subduced only the spatial components of the current\footnote{Use of an anisotropic lattice requires us to treat the spatial and temporal components of the current separately, as we will discuss later.}.

\subsection{Correlator and matrix-element averaging procedure}
\label{WE-avg}

The set of all possible correlation functions of the type presented in Eqn.~\ref{eq:3ptIrrepLabels}, even when restricted to the limited set of vector current insertion momentum direction $\mathbf{q}$ mentioned earlier, will include many which are related by allowed lattice rotations of the initial and final momenta, or combinations which contain different rows ($\mu$) of the same irrep ($\Lambda$).
We can make use of the symmetries of the (boosted) lattice to relate many of the operator combinations to each other, and specifically we will use the group-theoretic Wigner-Eckart theorem which takes advantage of Clebsch-Gordan coefficients for the relevant symmetry groups. We'll follow the notation laid out in Ref.~\cite{Dudek:2012gj} to express application of the theorem as
\begin{align}
\label{eq:WETheorem}
 \big\langle 0 \big|  \, 
  \Omega&_{\eta^{(\prime)}}^{\{\mathbf{p}'\}^\star_{k'} \Lambda'\mu'} 
  \,\cdot j^{\mathbf{q} \Lambda_{\gamma} \mu_{\gamma}}
  \,\cdot \left[\Omega_\psi^{\{\mathbf{p}\}^\star_k\Lambda \mu}\right]^\dag\!\! \,  
  \big|0 \big\rangle \nonumber\\
  &= \mathcal{C}\big(    \{ \mathbf{p} \}^{\star}_k \Lambda \mu;\,  \{ \mathbf{p}' \}^{\star}_{k'}\Lambda' \mu' ;\;  \mathbf{q}\, \Lambda_{\gamma} \mu_{\gamma} \big ) \nonumber\\
  &\quad\quad\quad\quad\quad\quad \cdot \mathbb{C}
  \big(\{\mathbf{p}'\}^\star\Lambda'; \mathbf{q}\Lambda_\gamma; \{\mathbf{p}\}^\star\Lambda \big) \,.
\end{align}
The \emph{star} of a momentum, $\{ \mathbf{p}\, \}^{\star}$, is the set of all momenta related to $\mathbf{p}$ by an allowed lattice rotation, with the subscript $k$ labeling the particular direction of the momentum. The objects $\mathcal{C}$ are the Clebsch-Gordan coefficients for $\{ \mathbf{p}' \}^{\star} \Lambda' \otimes \mathbf{q}\Lambda_\gamma \to \{ \mathbf{p} \}^{\star}\Lambda$, for momenta such that $\mathbf{p}' + \mathbf{q} = \mathbf{p}$.
The object $\mathbb{C}$ 
is the \emph{reduced correlation function} which no longer carries any dependence upon the momentum directions or the irrep rows\footnote{In some cases $\{ \mathbf{p}' \}^{\star} \Lambda' \otimes \mathbf{q}\Lambda_\gamma$ can combine to $\{ \mathbf{p} \}^{\star}\Lambda$ in more than one way, with each referred to as an \emph{embedding}. Our operator constructions are such that in these cases we have independent three--point functions leading to separate reduced correlation functions for each embedding.}.

The Clebsch-Gordan coefficients are calculated following the method outlined in Ref.~\cite{Dudek:2012gj}, except that the direction of the momentum of the current insertion, $\mathbf{q}$, is held fixed and all possible lattice rotations are applied to the initial and final momenta.

In practice we obtain the reduced correlation function from each related correlation function by dividing out the Clebsch-Gordan coefficient, and average this set, before multiplying back in the Clebsch-Gordan coefficient for a single reference case. After applying this method we are left with one correlation function for each embedding of the combination of irreps in the matrix element, and this can then be subject to a timeslice fit, yielding a single matrix--element value. Further details, and some illustrative examples, are presented in Appendix~\ref{CGs}.

\medskip

A second stage of averaging can be performed, one which is not a reflection of an exact lattice symmetry, but which rather relies upon there being modest discretization effects in, for example, the subduction of the vector current into different irreps\footnote{In Appendix~\ref{discretization} we show an example which supports this assumption.}. The matrix--element decomposition presented in Eqn.~\ref{eq:K} assumes Lorentz symmetry, and when subduced in a manner consistent with our operator constructions, yields
\begin{align*}
  \begin{split}
  &\langle \eta^{(\prime)}_{\Lambda'}(\mathbf{p}') | j^{\mathbf{q}\Lambda_{\gamma} \mu_{\gamma}} | \psi_{\Lambda \mu}(\mathbf{p})\rangle \\
  &\quad =\left[\sum_{\hat{\lambda}_{\gamma}} S^{\tilde{\eta}_{\gamma}, \hat{\lambda}_{\gamma}}_{\Lambda_{\gamma}, \mu_{\gamma}}  \sum_m D^{(1)\,*}_{m, \hat{\lambda}_{\gamma}}(R_\mathbf{q}) (-i) {\epsilon}^{*}_{j}(\mathbf{0},m) \right] \\
  &\quad\qquad \times \left[ \sum_{\hat{\lambda}} S^{\tilde{\eta}_\psi, \hat{\lambda}}_{\Lambda, \mu} {\epsilon}_{k}(\mathbf{p}, \hat{\lambda}) \right] 
  \epsilon^{j\nu\rho k} p'_{\nu} p_{\rho} \,  F_{\psi \eta^{(\prime)}}(Q^2) \\
  &\quad = K(\mathbf{p'}\Lambda';\, \mathbf{q}\Lambda_\gamma \mu_\gamma; \, \mathbf{p}\Lambda\mu) \cdot  F_{\psi \eta^{(\prime)}}(Q^2)  \, ,
\end{split}
\end{align*}
where assuming a common form--factor value for the different irrep choices at a fixed $Q^2$ is equivalent to assuming an effective Lorentz symmetry.

\begin{figure}
  \begin{subfigure}[b]{\columnwidth}
    \centerline{\includegraphics[width=.92\columnwidth]{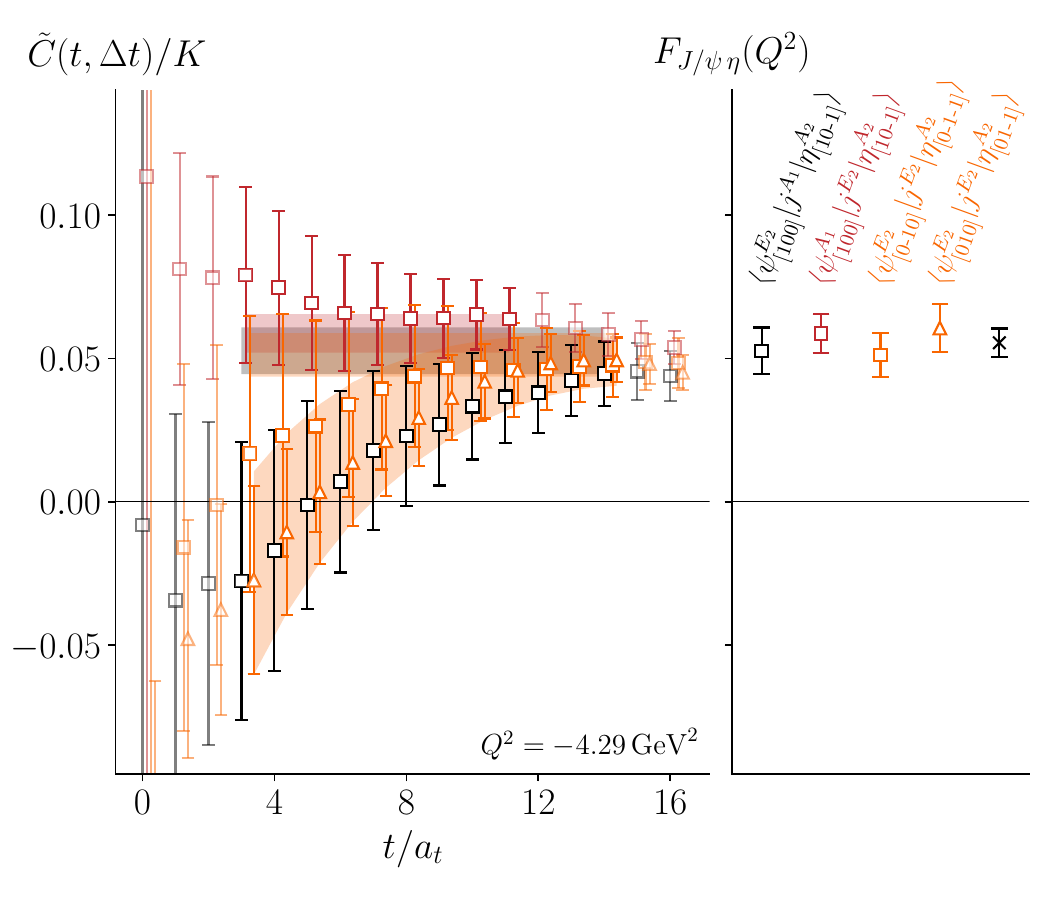}}
  \end{subfigure}
  \begin{subfigure}[b]{\columnwidth}
    \centerline{\includegraphics[width=.92\columnwidth]{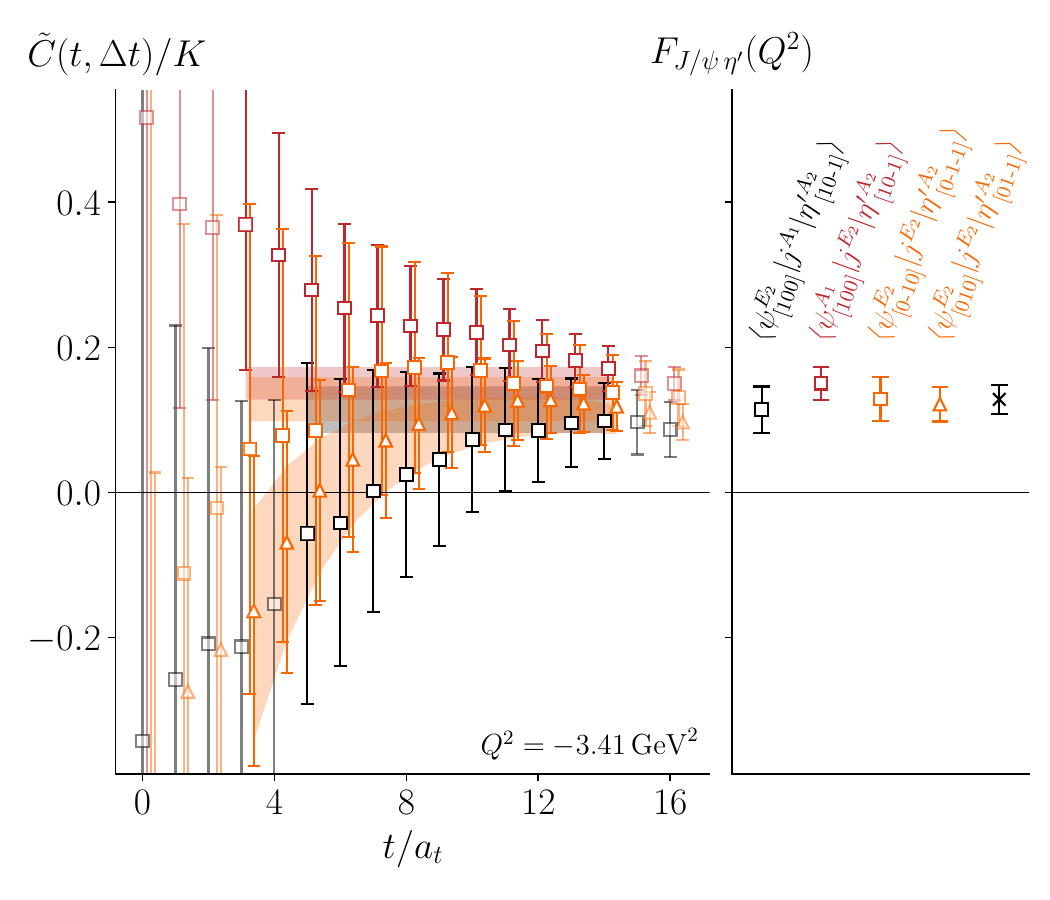}}
  \end{subfigure}
  \vspace*{-8mm}
  \caption[]{\label{fig:fit-solve} 
  Example of correlator averaging procedure for $J/\psi \to \gamma \eta$ (top) and $J/\psi \to \gamma \eta'$ (bottom) as described in the text. The rightmost point in the right panels shows the result of the complete set of averaging procedures.
  }
\end{figure}

We combine the matrix--elements determined from different irrep choices at a fixed $Q^2$ in an overconstrained linear least--squares manner (which generalizes to the case of several form--factors each with an independent kinematic factor) by computing
\begin{equation}
\label{eq:averaging}
F = (K^T \Sigma^{-1}K)^{-1}\, K^T\Sigma^{-1} \, \Gamma \,,
\end{equation}
where $\Gamma$ is a vector containing the various matrix--element values ($J$ in the notation used previously), and $\Sigma$ is the data covariance matrix for these, with $K$ being the vector of kinematic function values.

Figure~\ref{fig:fit-solve} illustrates this averaging approach for two examples, one describing $J/\psi \to \gamma \, \eta$ (top) and another $J/\psi \to \gamma \,\eta'$ (bottom). In each case, the left panel shows the result obtained by performing the Wigner-Eckart averaging, for four different irrep combinations or embeddings at a particular $Q^2$. The single timeslice fit with highest AIC--value is also shown for each case, illustrating the quality of description. The right panel shows the values of the form--factor extracted from timeslice fitting, presenting the values from AIC--averaging results obatined over many time-windows. The individual determinations of the form--factor from each three--point function are observed to be compatible, supporting the hypothesis that they can be averaged, and the result of doing so using Eqn.~\ref{eq:averaging} is shown as the rightmost point. This average can be seen to have a somewhat reduced uncertainty relative to any one of the averaged points.

\subsection{Vector Current Improvement and Renormalization}

\begin{figure}[b]
  \centerline{\includegraphics[width=.99\columnwidth]{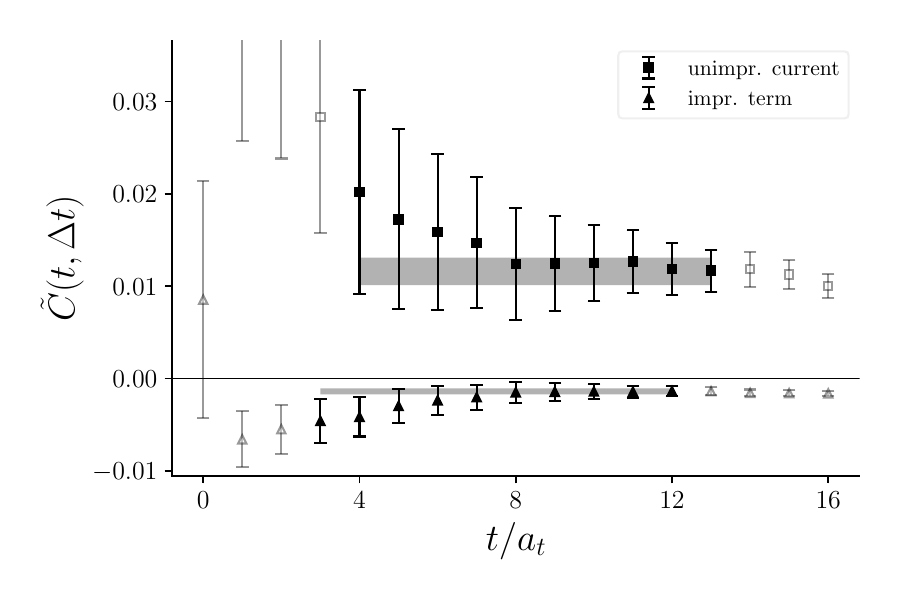}}
  \caption[]{\label{fig:impr-term} Contributions to a three-point function for $J/\psi \to\gamma\, \eta$ with $\mathbf{n}_{\eta} = [1,\text{-}1,\text{-}2]$, $\mathbf{n}_{\psi} = [1,1,1]$ where $Q^2 = 0.23 \,\textrm{GeV}^2$. For illustration the unimproved current and the improvement term have separate timeslice fits.
  }
\end{figure}

The local vector current we use in this calculation needs to be multiplicatively renormalized by a factor which can be determined by insisting that the $\eta_c$ form--factor take unit value at $Q^2=0$. For both spatially and temporally directed currents, these non-perturbative renormalization constants have been calculated on this lattice in Ref.~\cite{Delaney:2023fsc},
\begin{equation*}
\label{eq:23}
Z_V^s = 1.145(3), \qquad Z_V^t = 1.253(3).
\end{equation*}
In order to have a consistent Symanzik improvement on this anisotropic Clover lattice, we must also transform the quark fields in the vector current, leading us to include a \emph{tree--level} \mbox{$\mathcal{O}(a)$-improvement term}~\cite{Shultz:2015pfa},
\begin{align*}
  j_k &= Z_V^s \Big( \bar{\psi} \gamma_k \psi + \tfrac{1}{4} (1-\xi) a_t \partial_0 (\bar{\psi} \sigma_{0k} \psi ) \Big),\\
  j_0 &= Z_V^t \Big( \bar{\psi} \gamma_0 \psi + \tfrac{1}{4}\tfrac{\nu_s}{\xi} (1-\xi) a_s \partial_j (\bar{\psi} \sigma_{0j} \psi ) \Big) \, ,
\end{align*}
where $\nu_{s}$ is a parameter appearing in the fermion action taking value $\nu_{s} = 1.078$ for charm quarks, and ${\sigma_{\mu\nu} = \frac{1}{2}[\gamma_{\mu}, \gamma_{\nu}]}$.
For the anisotropy $\xi$ in the improvement factor, we will use the value determined from fits to the $J/\psi$ dispersion relation described in Section \ref{twoptfunctions}. The prefactor for the improvement term of the spatial current depends on the energy difference between the initial and final state and as such will vary in magnitude over the kinematic range we consider.
In Figure~\ref{fig:impr-term} we show the contributions to a typical three--point function (for a value of $Q^2$ close to 0) from the spatially directed unimproved vector current and from the improvement term, where it can be seen that in this case the improvement term contributes at the $15\%$ level.
%

In practice we will add the improvement term to the unimproved current before correlator averaging and timeslice fitting, thus determining the improved--current matrix element.

\subsection{Dataset for form--factor determination}

For the remainder of this paper we will work with a dataset constructed from all three--point correlation functions with $|\mathbf{n}_{\psi}|^2\le 4$, $|\mathbf{n}_{\eta^{(\prime)}}|^2\le 6$, and $|\mathbf{n}_{\mathbf{q}}|^2\le 16$, separately for $\eta$ and $\eta'$, with $\Delta t / a_t = 16$. In Appendix~\ref{discretization} we will present results extracted from correlators with ${\Delta t/a_t = 12, 20, 24}$ showing that larger source--sink separations lead to consistent results with larger statistical uncertainties.

In total we generated 1748 correlation functions, which reduce down to 70 values of $Q^2$ for each of $\eta$ and $\eta'$ upon application of the averaging procedure described above. The matrix--element values that we will present in the next section thus benefit from averaging over multiple time--sources and multiple rotationally equivalent kinematic configurations, while also having conservative uncertainties estimated from AIC--averaging over a range of time--window fits.


\section{Radiative Decay Form Factor}\label{formfactors}

\begin{figure*}
  \begin{subfigure}[b]{\textwidth}
    \centerline{\includegraphics[width=.8\textwidth]{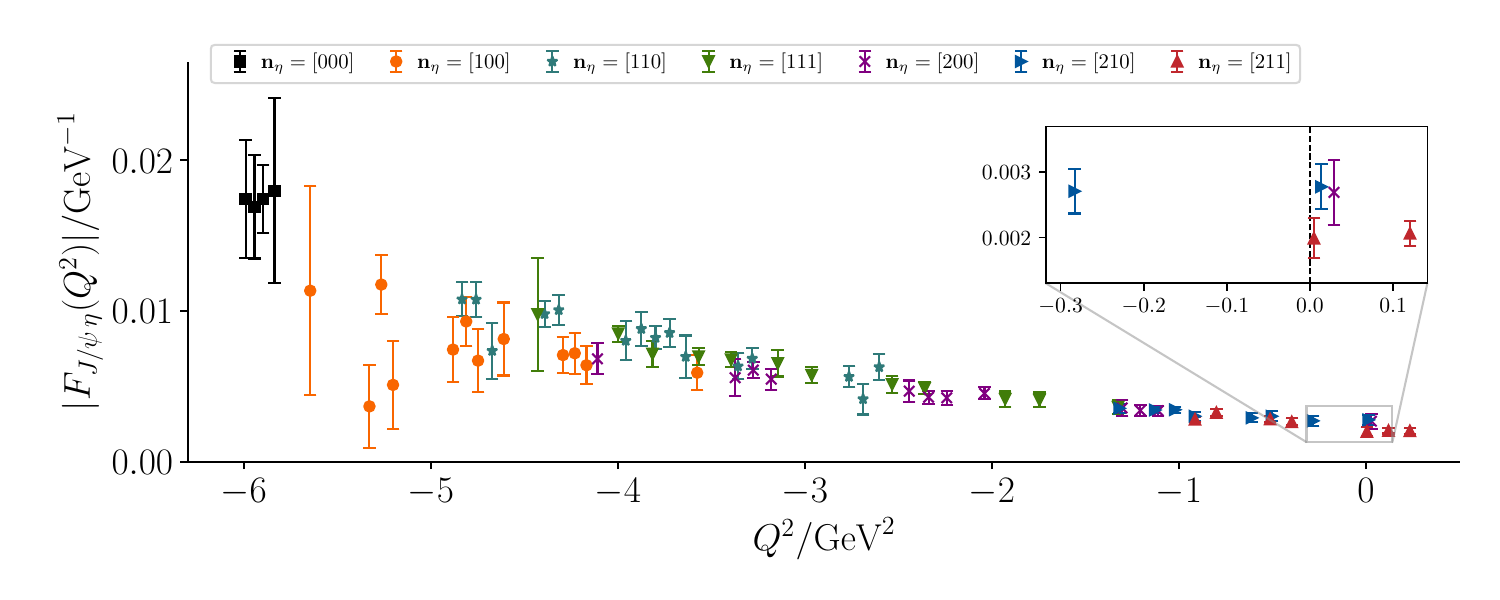}}
  \end{subfigure}
  \begin{subfigure}[b]{\textwidth}
    \centerline{\includegraphics[width=.8\textwidth]{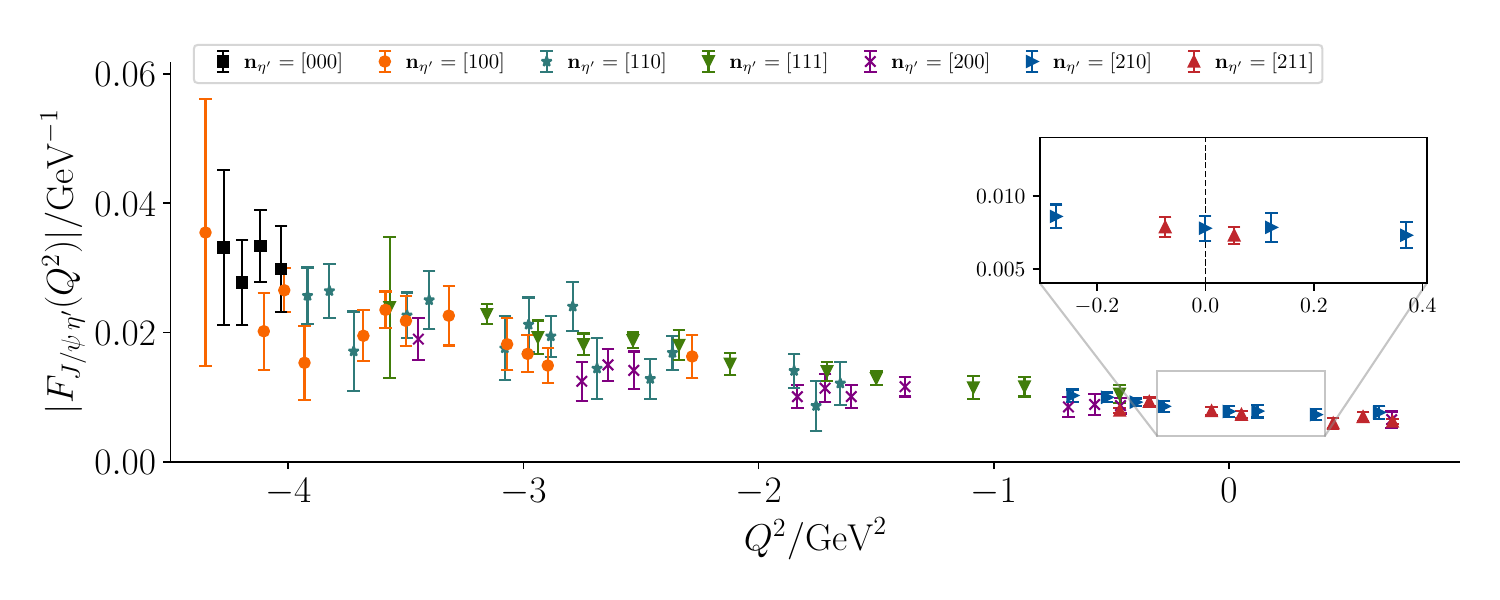}}
  \end{subfigure}
  \vspace*{-8mm}
  \caption[]{\label{fig:ff-snk-highlight} Form--factor for $J/\psi\to\gamma\, \eta$ (top) and $J/\psi\to\gamma \, \eta'$ (bottom) as a function of photon virtuality, $Q^2$. Insets show zoomed regions around the real--photon point which can be interpolated linearly for an estimate of $|F(0)|$.}
\end{figure*}

\begin{figure*}
  \centerline{\includegraphics[width=.9\textwidth]{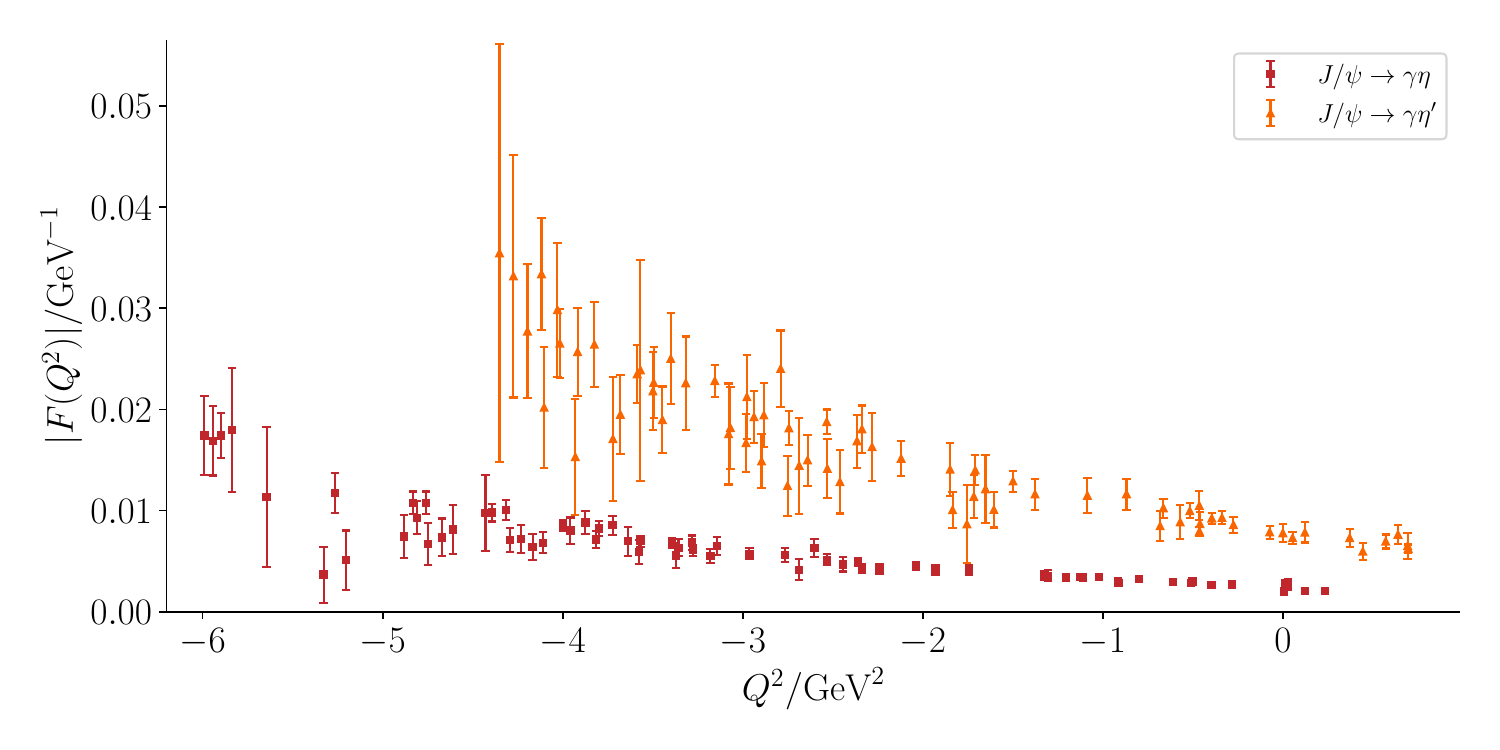}}
  \caption[]{\label{fig:ff-qsq}
    Form--factor for $J/\psi \to \gamma \,\eta$ and $J/\psi\to\gamma \,\eta'$ as a function of photon virtuality, $Q^2$.    
  }
\end{figure*}

Following the procedure described in the previous section, we obtain the form--factors $F_{J/\psi\, \eta}(Q^2)$ and $F_{J/\psi\, \eta^\prime}(Q^2)$ at 70 discrete points in photon virtuality. Figure~\ref{fig:ff-snk-highlight} shows the form--factor data colored according to the three--momentum of the $\eta^{(\prime)}$ operator used in the corresponding three--point functions. From this figure we can clearly see the necessity of utilizing high momentum to access the region near $Q^2=0$. For both the $\eta$ and $\eta'$ we observe (in general) good agreement between extractions from different three--momenta in common $Q^2$ regions. 
Figure~\ref{fig:ff-qsq} shows the two transitions together on the same scale. As expected, the dominantly $SU(3)_F$ singlet $\eta'$ has a significantly larger amplitude in this process, which proceeds through a gluonic intermediate state, than the dominantly octet $\eta$.

We have two objectives with this discretely sampled form--factor data: to determine the value at $Q^2=0$, corresponding to the true radiative process, and to describe the (mostly timelike) $Q^2$ behavior, which can be accessed experimentally in the Dalitz processes $J/\psi \to e^+ e^- \, \eta^{(\prime)}$. The $Q^2$ dependences will be described with a range of parameterization forms using correlated fits to the data plotted in Figure~\ref{fig:ff-qsq}.

\pagebreak
\subsection{$Q^2$--dependence of the form--factors}

Within the approximation we have made that the electromagnetic current couples only to the charm quark in the $J/\psi$, we do not expect any singularities in the region of $Q^2$ we have sampled. The nearest possible such singularity would be that due to a $J/\psi$ appearing in the off--shell photon, at $Q^2 = - m_{J/\psi}^2 \approx -9\,  \mathrm{GeV}^2$.

Motivated by the possible contributions of a vector meson, a common approach in the literature is to take a simple dipole form, often referred to as ``vector meson dominance'' (VMD),
\begin{equation*}
\label{eq:20}
F(Q^2) = \frac{F_0}{1 + Q^2/\Lambda^{2}} \, ,
\end{equation*}
which has a single pole singularity at $Q^2 = - \Lambda^2$.
It has been noted before~\cite{Dudek:2006ej} that there are good reasons why we should not expect this form to work well for charmonium, as the two nearest poles, the $J/\psi$ and $\psi(2S)$, are much closer together than the poles in the light sector, where VMD is usually applied\footnote{The argument extends to further excited unstable charmonium vector mesons which have poles off the real energy axis.}. Since the different poles can contribute to the form--factor with residues of differing sign, it is not clear what the scale of $\Lambda$ should be in a single--pole fit. 

\begin{figure}[htbp]
 \begin{subfigure}[b]{\columnwidth}
\centerline{\includegraphics[width=.99\columnwidth]{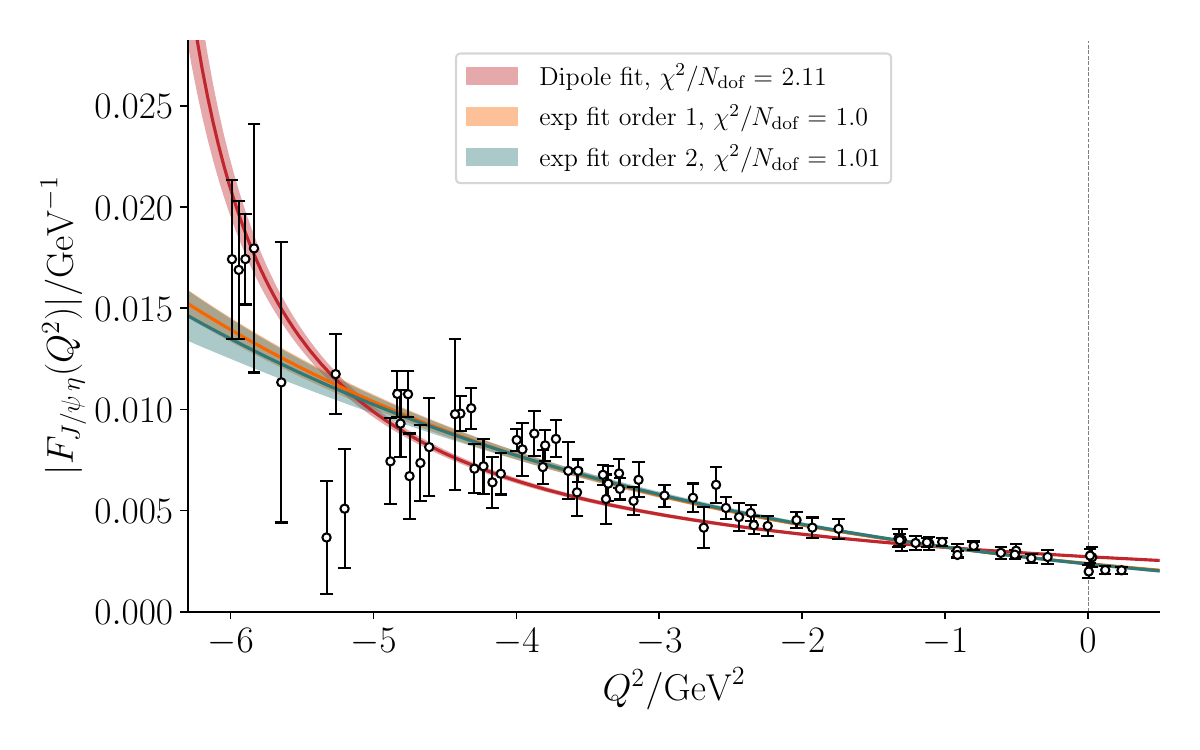}}
\end{subfigure}
\begin{subfigure}[b]{\columnwidth}
  \centerline{\includegraphics[width=.99\columnwidth]{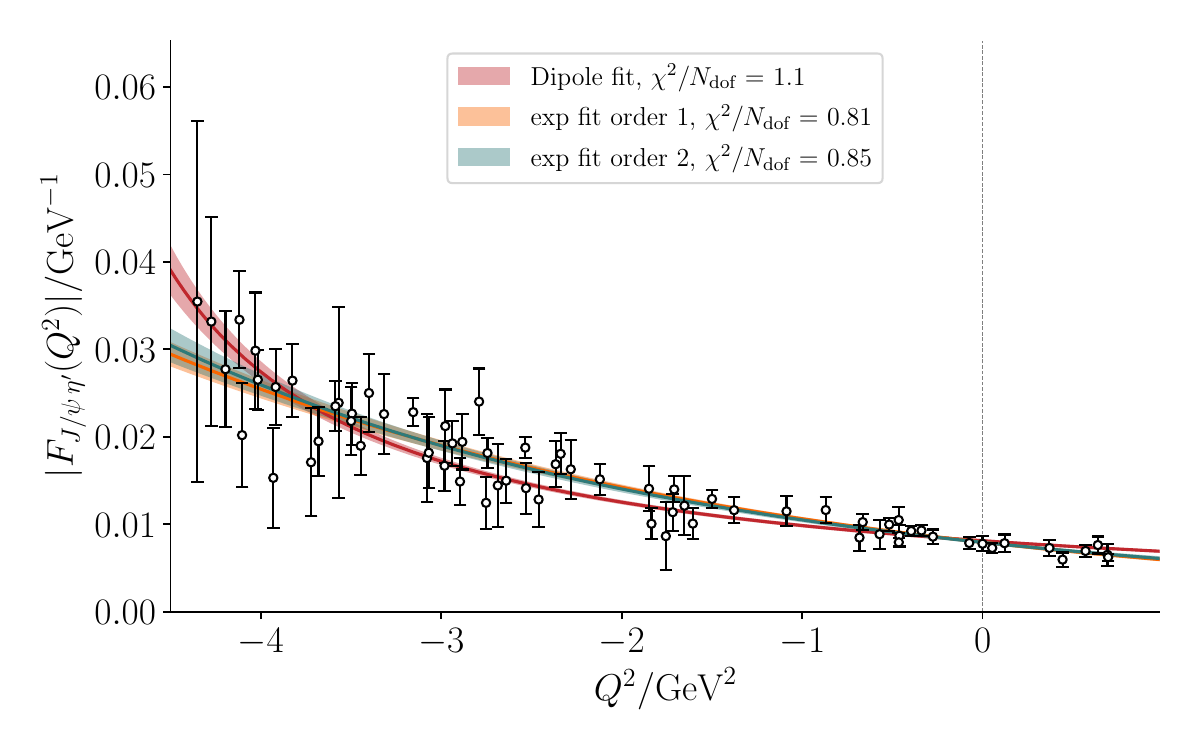}}
\end{subfigure}
\caption[]{\label{fig:qsq-fits-dipole-exp} Form--factors for $J/\psi\to\gamma\eta$ (upper) and $J/\psi\to\gamma\eta'$ (lower), showing results of dipole and exponential fits.}
\end{figure}

Figure~\ref{fig:qsq-fits-dipole-exp} includes the results of dipole fits to both form--factors. For the $J/\psi \to \gamma \, \eta$ case, where our data goes deeper into the timelike region, spanning a larger range in $Q^2$, this form does not describe the data well, with a $\chi^2/N_{\textrm{dof}}=2.11$, and a systematic trend visibly different to the data. The $J/\psi \to \gamma \, \eta'$ case is described better, but this is over a more limited $Q^2$ region.
The dipole scale parameters extracted from these fits are
\begin{equation*}
\label{eq:28}
\Lambda_{J/\psi\,\eta} = 2.596(14) \,\textrm{GeV}, \quad \Lambda_{J/\psi\,\eta'} = 2.356(25) \,\textrm{GeV} \, ,
\end{equation*}
which we note are significantly lower than the mass of the $J/\psi$. Notably, for the $J/\psi\to\gamma\, \eta$ case, the determined $\Lambda$ value is quite similar to the value extracted in a fit\footnote{Their fit--form also allows for a modest contribution from a light vector meson, which practically has a contribution only in a single $Q^2$ bin.} to the corresponding Dalitz decay data from BESIII~\cite{BESIII:2018qzg}, $\Lambda = 2.56(4) \,\textrm{GeV}$.

\medskip

A phenomenological $Q^2$ parameterization form~\cite{Dudek:2006ej} which features no nearby singularities is
\begin{equation*}
F(Q^2) = F_0\, e^{-Q^2/16\beta^2} \, ,
\end{equation*}
which can be made more flexible by adding an additional term in the exponential,
\begin{equation*}
F(Q^2) = F_0\, e^{-(1 + \alpha Q^2)Q^2/16\beta^2}.
\end{equation*}
Fits to these forms are also shown in Figure~\ref{fig:qsq-fits-dipole-exp}, where they are observed to give a better description of the $Q^2$ dependences than the dipole form.

\medskip

\begin{figure}[b]
 \begin{subfigure}[b]{\columnwidth}
\centerline{\includegraphics[width=.99\columnwidth]{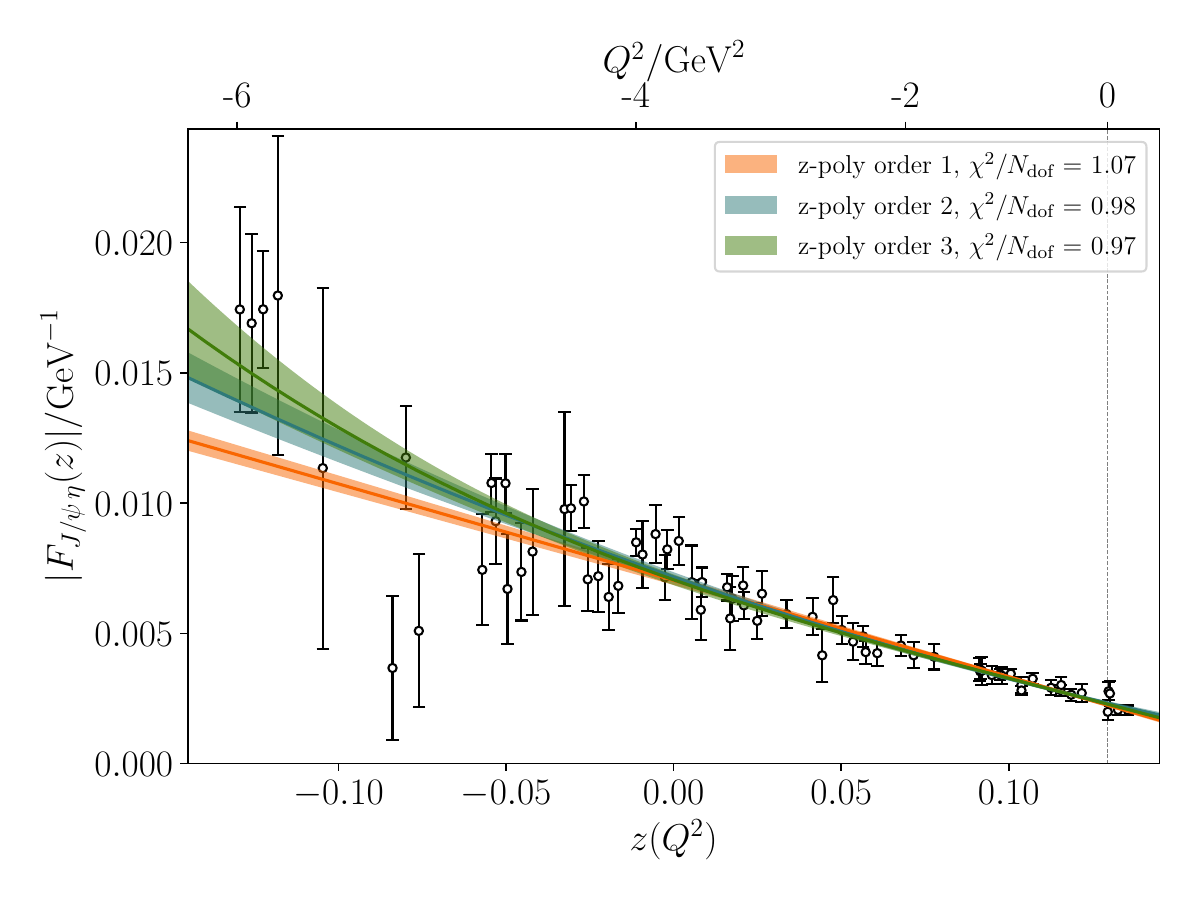}}
\end{subfigure}
  \begin{subfigure}[b]{\columnwidth}
\centerline{\includegraphics[width=.99\columnwidth]{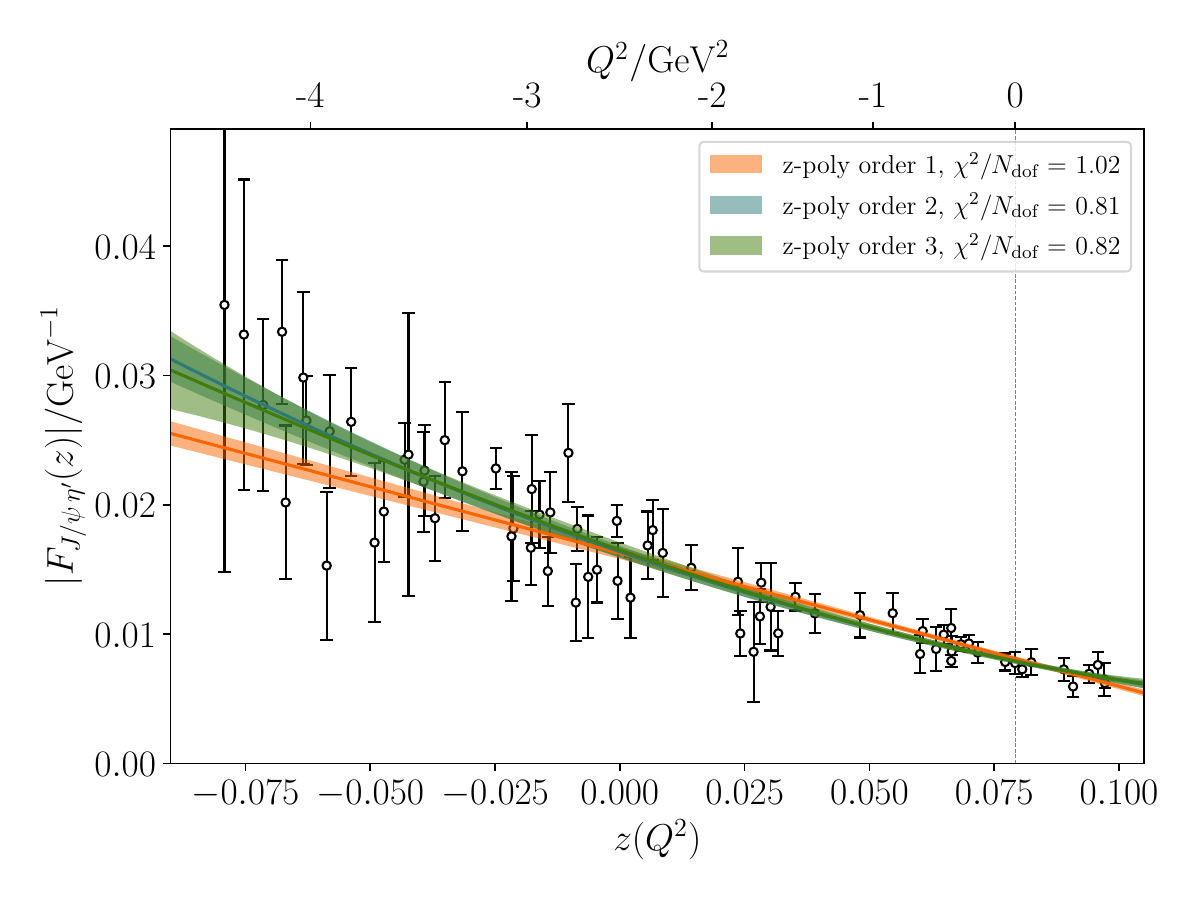}}
\end{subfigure}
\caption[]{\label{fig:qsq-fits-zexp} Form--factors for $J/\psi\to\gamma\, \eta$ (upper) and $J/\psi\to\gamma\, \eta'$ (lower), expressed in terms of $z$ defined by Eqn.~\ref{eq:z}, fitted with polynomial forms, Eqn.~\ref{eq:zpoly}. Vertical dashed line in each case corresponds to $Q^2=0$.}
\end{figure}

Another parameterization scheme makes use of the lack of nearby singularities to perform a conformal mapping of the $Q^2$ plane into a unit disk of a variable $z(Q^2)$, with a polynomial in $z$ expected to converge rapidly.  
The conformal variable is defined as
\begin{equation}
\label{eq:z}
z(Q^2, t_{\text{cut}}, t_{0}) = \frac{\sqrt{t_{\text{cut}} + Q^2} - \sqrt{t_{\text{cut}} - t_{0}}}{\sqrt{t_{\text{cut}} + Q^2} + \sqrt{t_{\text{cut}} - t_{0}}} \, ,
\end{equation}
where real $Q^2$ values more timelike than $-t_{\text{cut}}$ lie outside the unit circle in $z$. We choose to select $\sqrt{t_{\text{cut}}} = m_{J/\psi}$ so that the nearest singularity (which is relatively distant from our data) is excluded. The parameter $t_{0}$ sets the $-Q^2$ value which maps to $z = 0$, and can be chosen suitably to center the data\footnote{
Choosing $\sqrt{t_0} = 1.926\,\mathrm{GeV}, \, 1.587 \,\mathrm{GeV}$ for the $\eta,\, \eta'$ cases respectively achieves this.
}
around $z=0$.
Using this conformal mapping, we perform polynomial fits to the form--factor,
\begin{equation}
\label{eq:zpoly}
F(Q^2) = \sum_{n=0}^{k}a_n \, z(Q^2)^{n},
\end{equation}
where we consider polynomials for orders $k \le 3$. The resulting fits are shown in Figure~\ref{fig:qsq-fits-zexp} where it is clear that already at the $k=1$ level, the bulk of the $Q^2$ dependence of the data is captured. Making modest adjustments to the location of $t_\mathrm{cut}$ and $t_0$ does not significantly impact the quality of the data description.

\smallskip
Details of the parameterizations plotted in Figures~\ref{fig:qsq-fits-dipole-exp},~\ref{fig:qsq-fits-zexp} are given in Appendix \ref{qsq-fits}.

\pagebreak
\subsection{Radiative decay widths}

The radiative partial decay widths are given by Eqn.~\ref{eq:width}, which requires knowledge of the form--factor at ${Q^2=0}$. We can estimate $F(0)$ by evaluating our previous \mbox{form--factor} parameterizations at $Q^2=0$, or we can make use of just our discrete determinations in the region around $Q^2=0$ as shown in the insets in Figure~\ref{fig:ff-snk-highlight}, by fitting this limited set of data as a straight line in $Q^2$. These linear fits yield values
\begin{align*}
  |F_{J/\psi\,\eta}(0)| &= 0.00238(13) \,\textrm{GeV}^{-1}, \\
  |F_{J/\psi\,\eta'}(0)| &= 0.00786(33) \,\textrm{GeV}^{-1} \, ,
\end{align*}
while describing the small set of data well.

\begin{figure}
\centerline{\includegraphics[width=.97\columnwidth]{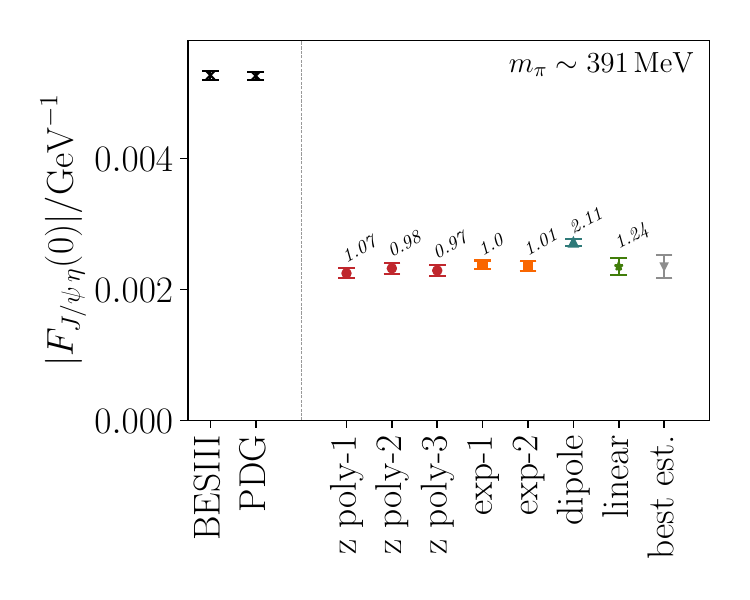}}
\caption[]{\label{fig:exp-comparison-jpsi-eta} Form--factor for $J/\psi\to\gamma\,\eta$ on this paper's lattice from parameterizations as discussed in the text. $\chi^2/N_\mathrm{dof}$ for each fit shown in italics.
Also shown are extractions from the PDG average partial decay width, and the most precise single measurement from BESIII~\cite{BESIII:2023fai} (which is included in the PDG average). The grey point shows our conservative best estimate on this lattice.}
\end{figure}

\begin{figure}
\centerline{\includegraphics[width=.97\columnwidth]{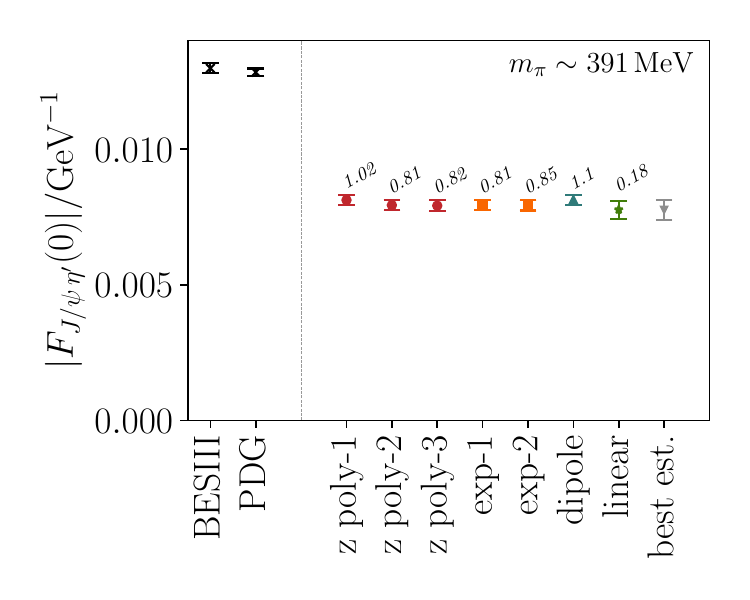}}
\caption[]{\label{fig:exp-comparison-jpsi-eta-prime} Form--factor for $J/\psi\to\gamma\,\eta'$ on this paper's lattice from parameterizations as discussed in the text. $\chi^2/N_\mathrm{dof}$ for each fit shown in italics.
Also shown are extractions from the PDG average partial decay width, and the most precise single measurement from BESIII~\cite{BESIII:2023fai} (which is included in the PDG average). The grey point shows our conservative best estimate on this lattice.}
\end{figure}

In Figures~\ref{fig:exp-comparison-jpsi-eta} and \ref{fig:exp-comparison-jpsi-eta-prime} we summarize our estimates of $F(0)$ for $J/\psi \to \gamma \, \eta$ and $J/\psi \to \gamma \, \eta'$ respectively, where we observe excellent agreement between different parameterization choices, with the dipole form for the $\eta$ being the only discrepancy (and as discussed in the previous section, this is the only parameterization which provides a poor description of the $Q^2$ dependence).
In Appendix \ref{discretization} we present studies of systematic variations in the analysis that justify a modest increase in uncertainty on our estimates of the form--factors, leading to our best estimates at $m_\pi \sim 391 \, \mathrm{MeV}$ of
\begin{align}
  \label{FINAL}
\begin{split}
  |F_{J/\psi\,\eta}(0)| &= 0.00235(18) \,\textrm{GeV}^{-1}, \\
  |F_{J/\psi\,\eta'}(0)| &= 0.00777(37) \,\textrm{GeV}^{-1} \, .
\end{split}
\end{align} 
These values are shown as the gray triangles on the far right in Figures~\ref{fig:exp-comparison-jpsi-eta} and \ref{fig:exp-comparison-jpsi-eta-prime}.
Using these best estimates, the ratio of form--factors is estimated to be
\begin{equation}
\label{eq:17}
\frac{|F_{J/\psi\,\eta'}(0)|}{|F_{J/\psi\,\eta}(0)|} = 3.30(29) \, ,
\end{equation}
which is observed to be somewhat larger than the value inferred from PDG--averaged branching fractions,
\begin{equation}
\label{eq:16}
\frac{|F_{J/\psi\,\eta'}(0)|}{|F_{J/\psi\,\eta}(0)|} = 2.44(4) \, .
\end{equation}
%

\medskip

The light--quark mass dependence of these form--factors is not known to us, and as such it may be the case that the determined values indeed evolve to the experimental values as the light quark mass is reduced to its physical value. Some possible guidance comes from Ref.~\cite{Novikov:1979uy}, where $SU(3)_F$ symmetry and the Adler--Bell--Jackiw anomaly are used to argue that the gluonic matrix element leading to production of the $\eta$ is proportional to $f_\pi m_\eta^2$, while nothing else in the radiative decay is expected to be sensitive to the light quark mass. The quark mass evolution of the $\eta$ mass and $f_\pi$ (or $f_{\eta_8}$ which would be an acceptable replacement in the limit of $SU(3)_F$ symmetry) would not be expected to be sufficient to bring our result into agreement with experiment. The non-Goldstone nature of the $\eta'$ required a more involved and assumption--dependent sum--rule application in Ref.~\cite{Novikov:1979uy}, but again there was nothing in the study that would indicate particularly strong light--quark mass dependence. The methodology of Ref.~\cite{Novikov:1979uy} underestimates the partial widths $\Gamma(J/\psi \to \gamma \, \eta^{(\prime)})$ compared to the modern PDG averages, with the $\eta'$ rate being particularly discrepant.


\section{Conclusion}\label{conclusion}

As discussed in the previous section, our calculation at $m_\pi \sim 391$ MeV yields transition rates for $J/\psi \to \gamma \, \eta$ and $J/\psi \to \gamma \, \eta'$ that are significantly lower than experimental determinations, while the ratio of the rates is in the ballpark of experiment, but still somewhat discrepant. Absent computation at other light--quark masses, we cannot eliminate the possibility that these observables are particularly sensitive to the light--quark mass, but we can also consider other possible sources of discrepancy.

One possible source would be lattice spacing dependence, which we cannot directly estimate with a calculation on a single lattice spacing. We carried out the Symanzik improvement of the heavy quarks consistently, including in the vector current, and as observed in Figure~\ref{fig:decay-const} in Appendix~\ref{decay-const}, the implemented improvement term appears to reduce irrep--dependence and bring the spatially--directed current estimates of the $J/\psi$ decay constant into reasonable agreement with the experimental value. We might tentatively suggest that the possible $\mathcal{O}(a)$ discretization effects have been removed, but there remains the important observation that even with this improved action, the $J/\psi$--$\eta_c$ mass splitting is underpredicted significantly~\cite{HadronSpectrum:2012gic}. 

Another possibility might be that the particular ensemble of gauge field configurations used in this calculation might not adequately reflect the true topology of the QCD vacuum. We recall that the $\eta'$ mass determined on this lattice is found to be lighter than we might expect given the large light--quark mass, and a possible explanation would come if the $U(1)_A$ anomaly contribution, which is sensitive to topology, is smaller than it should be. Figure 12 in Ref.~\cite{Dudek:2013yja} shows an unexpectedly large volume dependence in the mass of the $\eta'$, considering that it is QCD--stable, and this remains unexplained.
The radiative decays considered in this paper may be particularly sensitive to any flawed lattice topology, even for the $\eta$, because only the $SU(3)_F$ singlet component of the pseudoscalar mesons is populated in the $c\bar{c}$ annihilation.

The only previous lattice QCD calculations of these radiative decays, presented in Refs.~\cite{Jiang:2022gnd, Shi:2024fyv}, used lattices with either one or two flavors of light quark. To obtain reasonable signals, rather than the averaging techniques pursued in the current paper, they opted to use large numbers of configurations (several thousand compared to the several hundred used in the current calculation). They utilized a rather limited basis of meson interpolating operators, and hence did not benefit from the rapid relaxation afforded to us by the use of optimized operators, and they considered only a very limited set of kinematic configurations. Despite using an anisotropic Clover--improved action for the charm quarks, the corresponding Symanzik--program field transformation was not applied to the charm--quark fields in the vector current, leading to no implementation of the expected tree--level improvement term. This may conceivably lead to an unestimated $\mathcal{O}(a)$ error on the transition matrix elements, and this may be a significant effect.
Their extracted form--factor values, ${|F(0)|_{N_F=1} = 0.0054(1)\, \mathrm{GeV}^{-1}}$, ${|F(0)|_{N_F=2} = 0.0106(4)\, \mathrm{GeV}^{-1}}$, were obtained from $Q^2$ fits using a dipole form (which we found in our calculation may not be capable of describing the $\eta$ production process across a wide $Q^2$ region). 
Without having three flavors of quark, there cannot be a true distinction between the $\eta$ and the $\eta'$, and instead the authors of Refs.~\cite{Jiang:2022gnd, Shi:2024fyv} made phenomenological arguments to separate the two amplitudes from a single computed transition form--factor.

\medskip

Through use of optimized operators and novel averaging procedures, we've shown that high quality signals for matrix elements describing $J/\psi \to \gamma \, \eta$ and  $J/\psi \to \gamma \, \eta'$ can be obtained over a large kinematic region, from the deeply timelike through the real photon point. Given the observed discrepancy between the extracted transition form--factors and those extracted from experimental data, we propose applying the techniques demonstrated in this paper on other lattice configurations to explore the light--quark mass dependence, lattice spacing dependence, and to investigate the effect of vacuum topology by using lattices on which the topology is well-studied.

Going beyond the $J/\psi$, it would be relatively straightforward to extend to decays $h_c \to \gamma \eta^{(\prime)}$, since as seen in Figure~\ref{fig:disp-psi}, even when it appears as an excited state, the $h_c$ is well determined, and can be accessed using an appropriately constructed optimized operator. In this case we would be considering a $1^+ \to 0^-$ radiative transition, which has \emph{two} form--factors, so a slight modification in technology would be required to separate these~\cite{Dudek:2006ej, Dudek:2009kk}. Similarly the transitions $\psi' \to \gamma \, \eta^{(\prime)}$ can be studied using an optimized operator for the $\psi'$, and the obtained information may shed some light on the experimental mystery surrounding $\frac{\mathcal{B}(\psi(2S)\to\gamma\eta)}{\mathcal{B}(\psi(2S)\to\gamma\eta')}$.

The quality of the signals reported in this paper motivates computation of $J/\psi \to \gamma\,  \pi \pi$, $J/\psi \to \gamma \, K \bar{K} $ and other processes in which $f_J$ resonances can appear~\cite{Briceno:2017qmb}. The analysis of these will require the addition of the finite--volume two--hadron scattering formalism described in Refs.~\cite{Briceno:2021xlc, Briceno:2017max}(and references therein) and applied to other processes in Refs.~\cite{Briceno:2015dca, Briceno:2016kkp,  Radhakrishnan:2022ubg, Ortega-Gama:2024rqx, Alexandrou:2018jbt, Leskovec:2025gsw}. Demonstration of success there will justify extension to consider the exotic hybrid meson~\cite{Dudek:2011bn, Woss:2020ayi} candidate $\eta_1(1855)$ in $J/\psi \to \gamma \, \eta \eta'$.


\begin{acknowledgments}
We thank our colleagues within the Hadron Spectrum Collaboration for their continued assistance, and give particular thanks to C.E.~Thomas for his inspiration to consider the Wigner-Eckart averaging procedure. The authors acknowledge support from the U.S. Department of Energy contract DE-SC0018416 at William \& Mary, and contract DE-AC05-06OR23177, under which Jefferson Science Associates, LLC, manages and operates Jefferson Lab. 
This work contributes to the goals of the U.S. Department of Energy \emph{ExoHad} Topical Collaboration, Contract No. DE-SC0023598.
The authors acknowledge support from the U.S. Department of Energy, Office of Science, Office of Advanced Scientific Computing Research and Office of Nuclear Physics, Scientific Discovery through Advanced Computing (SciDAC) program. 
Also acknowledged is support from the Exascale Computing Project (17-SC-20-SC), a collaborative effort of the U.S. Department of Energy Office of Science and the National Nuclear Security Administration.

This work used clusters at Jefferson Laboratory under the USQCD Initiative and the LQCD ARRA project. This work also used the Cambridge Service for Data Driven Discovery (CSD3), part of which is operated by the University of Cambridge Research Computing Service (www.csd3.cam.ac.uk) on behalf of the STFC DiRAC HPC Facility (www.dirac.ac.uk). The DiRAC component of CSD3 was funded by BEIS capital funding via STFC capital grants ST/P002307/1 and ST/R002452/1 and STFC operations grant ST/R00689X/1. Other components were provided by Dell EMC and Intel using Tier-2 funding from the Engineering and Physical Sciences Research Council (capital grant EP/P020259/1). 

Also used was an award of computer time provided by the U.S.\ Department of Energy INCITE program and supported in part under an ALCC award, and resources at: the Oak Ridge Leadership Computing Facility, which is a DOE Office of Science User Facility supported under Contract DE-AC05-00OR22725; the National Energy Research Scientific Computing Center (NERSC), a U.S.\ Department of Energy Office of Science User Facility located at Lawrence Berkeley National Laboratory, operated under Contract No. DE-AC02-05CH11231; the Texas Advanced Computing Center (TACC) at The University of Texas at Austin; the Extreme Science and Engineering Discovery Environment (XSEDE), which is supported by National Science Foundation Grant No. ACI-1548562; and part of the Blue Waters sustained-petascale computing project, which is supported by the National Science Foundation (awards OCI-0725070 and ACI-1238993) and the state of Illinois. Blue Waters is a joint effort of the University of Illinois at Urbana-Champaign and its National Center for Supercomputing Applications.

The software codes
{\tt Chroma}~\cite{Edwards:2004sx}, {\tt QUDA}~\cite{Clark:2009wm,Babich:2010mu}, {\tt QUDA-MG}~\cite{Clark:SC2016}, {\tt QPhiX}~\cite{ISC13Phi},
{\tt MG\_PROTO}~\cite{MGProtoDownload}, and {\tt QOPQDP}~\cite{Osborn:2010mb,Babich:2010qb} were used.

\end{acknowledgments}

\bibliography{Library}

\appendix
\pagebreak

\section{$J/\psi$ Decay Constant}\label{decay-const}

\begin{figure}[b]
  \centerline{\includegraphics[width=.9\columnwidth]{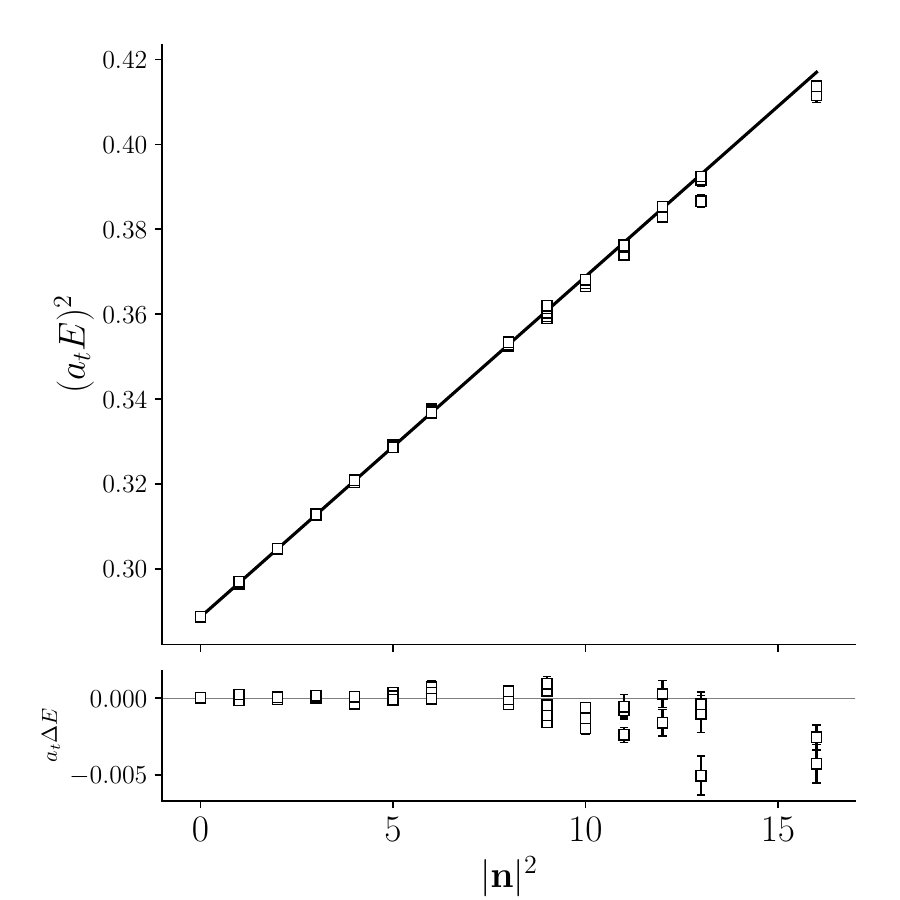}}
  \caption[]{\label{fig:disp-16} Dispersion relation of $J/\psi$, as in Figure~\ref{fig:disp-psi}, but here for momenta $|\mathbf{n}|^2\le 16$.
  }
\end{figure}

The charm--quark generalized perambulators computed in order to calculate $J/\psi \to \gamma \, \eta^{(\prime)}$ can also be used in a computation of the $J/\psi$ vector decay constant. This quantity can be defined in terms of the vector current matrix--element,
\begin{equation}
\label{eq:34}
  \big\langle 0 \big| j^{\mu}(0) \big| J/\psi(\mathbf{p},\lambda) \big\rangle = f_{\psi} m_{\psi} \, \epsilon^{\mu}(\mathbf{p},\lambda) \,,
\end{equation}
and this matrix--element appears in the two--point function,
\begin{equation*}
  \big\langle 0 \big| \Omega_\psi(t)\, j^{\mu}(0) \big|0 \big\rangle \, ,
\end{equation*}
using the optimized $J/\psi$ operator at momentum $\mathbf{p}$, and the renormalized spatial or temporal vector current $j^{\mu}$.
This correlator essentially corresponds to the left--half of Figure~\ref{fig:tikz-psice-gamma-eta}.
We computed the subductions of this two--point function considering each irrep at each momentum up to $\mathbf{p} = \left(\tfrac{2\pi}{L} \right)[0,0,4]$.
A ratio of this correlation function with the two--point function of the optimized operator was constructed to eliminate the ground--state energy dependence,
\begin{equation}
\label{eq:4}
R(t) = \frac{ \big\langle 0 \big| \Omega_\psi(t)\, j^{\mu}(0) \big|0 \big\rangle}{ \big\langle 0 \big| \Omega^{}_\psi(t)\, \Omega_\psi^{\dagger}(0) \big|0 \big\rangle} \, ,
\end{equation}
and this ratio was fitted to a constant or a constant plus an exponential, over a range of time-windows, and via an AIC--average, a value of $f_\psi$ was obtained.

Optimized operators for the $J/\psi$ were computed for higher momenta using the same technique described in the text, with the resulting dispersion relation being presented in Figure~\ref{fig:disp-16}. We observe that noticeable systematic departures from the relativistic dispersion relation are not seen until $|\mathbf{n}|^2 \gtrsim 10$, and that even then they are modest in size in absolute terms.

\begin{figure}[b]
 \begin{subfigure}[b]{\columnwidth}
\centerline{\includegraphics[width=.95\columnwidth]{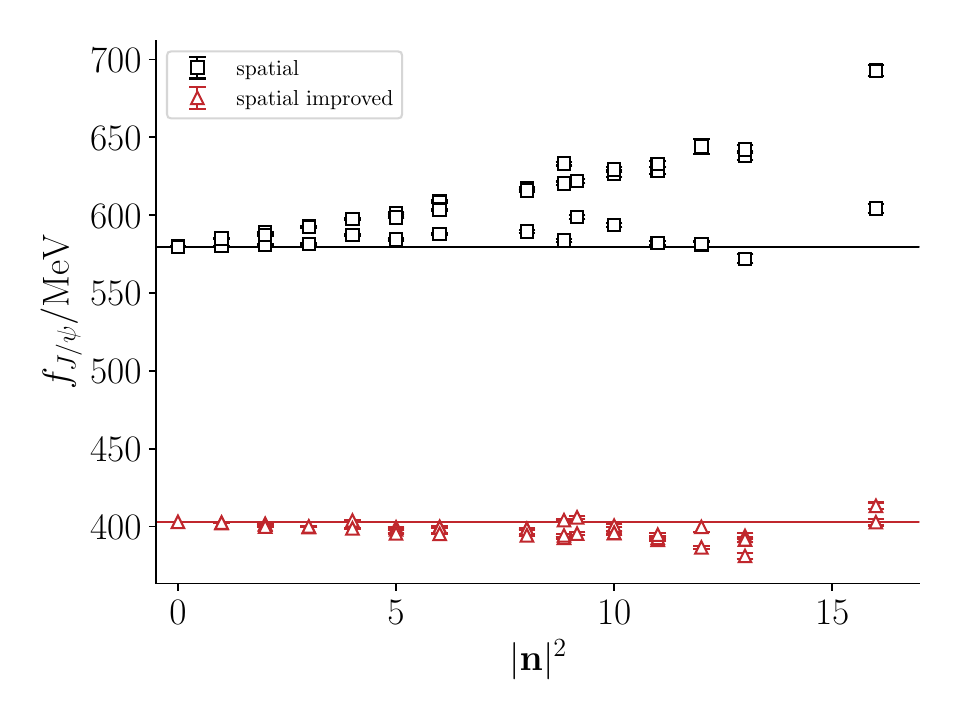}}
\vspace*{-4mm}
\end{subfigure}
\begin{subfigure}[b]{\columnwidth}
\centerline{\includegraphics[width=.95\columnwidth]{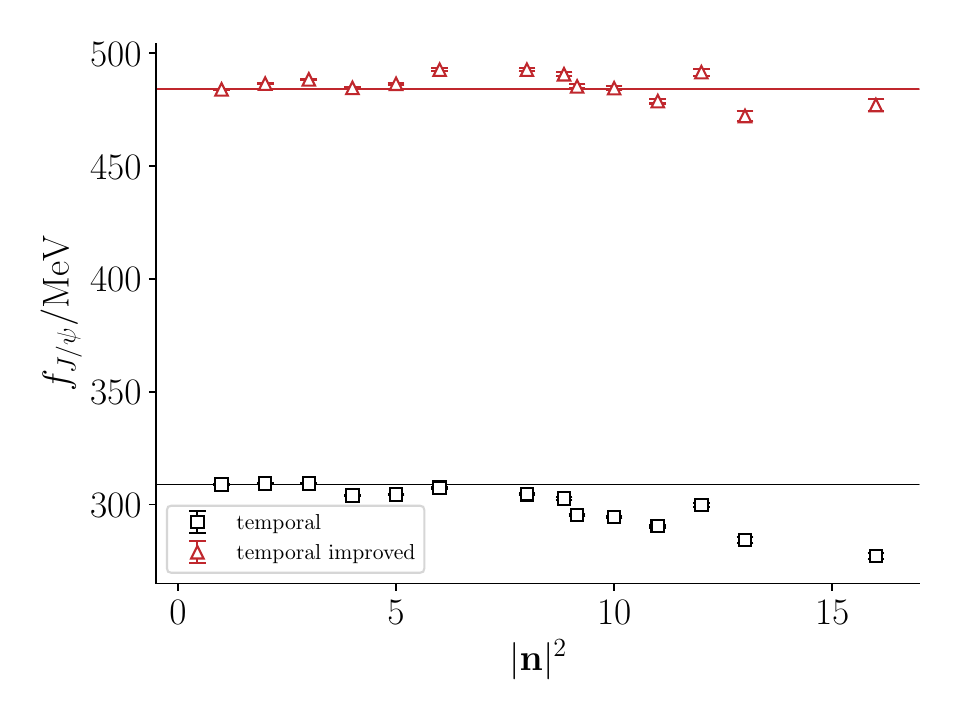}}
\end{subfigure}
\vspace*{-8mm}
\caption[]{\label{fig:decay-const} Decay constant of the $J/\psi$ using spatial (top) and temporal (bottom) currents for both the unimproved and improved vector current. Horizontal lines show values at $\mathbf{n} = [000]$ to guide the eye.}
\end{figure}

In Figure~\ref{fig:decay-const} we present $f_\psi$ for both spatial and temporal vector currents, with unimproved and improved vector currents shown for both. The absolute size of the improvement term is much larger here, relative to the radiative decays, since it is proportional to the energy difference between the vacuum and the energy of the $J/\psi$. We observe that for the spatial current, the addition of the improvement term removes almost all of the unwanted momentum dependence visible for the unimproved current, and significantly reduces discrepancies between irreps (corresponding to helicity $\lambda=0$ and $|\lambda|=1$) at fixed nonzero momentum. The remaining departures from constant behavior are at the few percent level, suggesting only modest discretization effects here.

The lower panel of Figure~\ref{fig:decay-const} shows that for the temporal current more significant momentum dependence is present even after improvement, and that there is a notable discrepancy between the magnitude of $f_\psi$ obtained from the spatial and temporal currents.

\medskip

The experimental value of the decay constant can be extracted from the leptonic decay width
\begin{equation}
\label{eq:15}
\Gamma(J/\psi \to e^+ e^- ) = \frac{4\pi}{3}\frac{4}{9} \alpha^2 \frac{f_{J/\psi}^2}{m_{J/\psi}} \, ,
\end{equation}
and using the current PDG values for the decay width \cite{ParticleDataGroup:2020ssz} this gives $|f_{J/\psi}| = 415(5)\,\textrm{MeV}$.
To compare this to the determination of the decay constant on this lattice we take the value from the improved spatial current at rest, $|f_{J/\psi}| = 402.99(16)\,\textrm{MeV}$, which appears to be in agreement with experiment at the few percent level.


\section{Averaging Procedure}\label{CGs}

\begin{figure*}
  \centerline{\includegraphics[width=.95\textwidth]{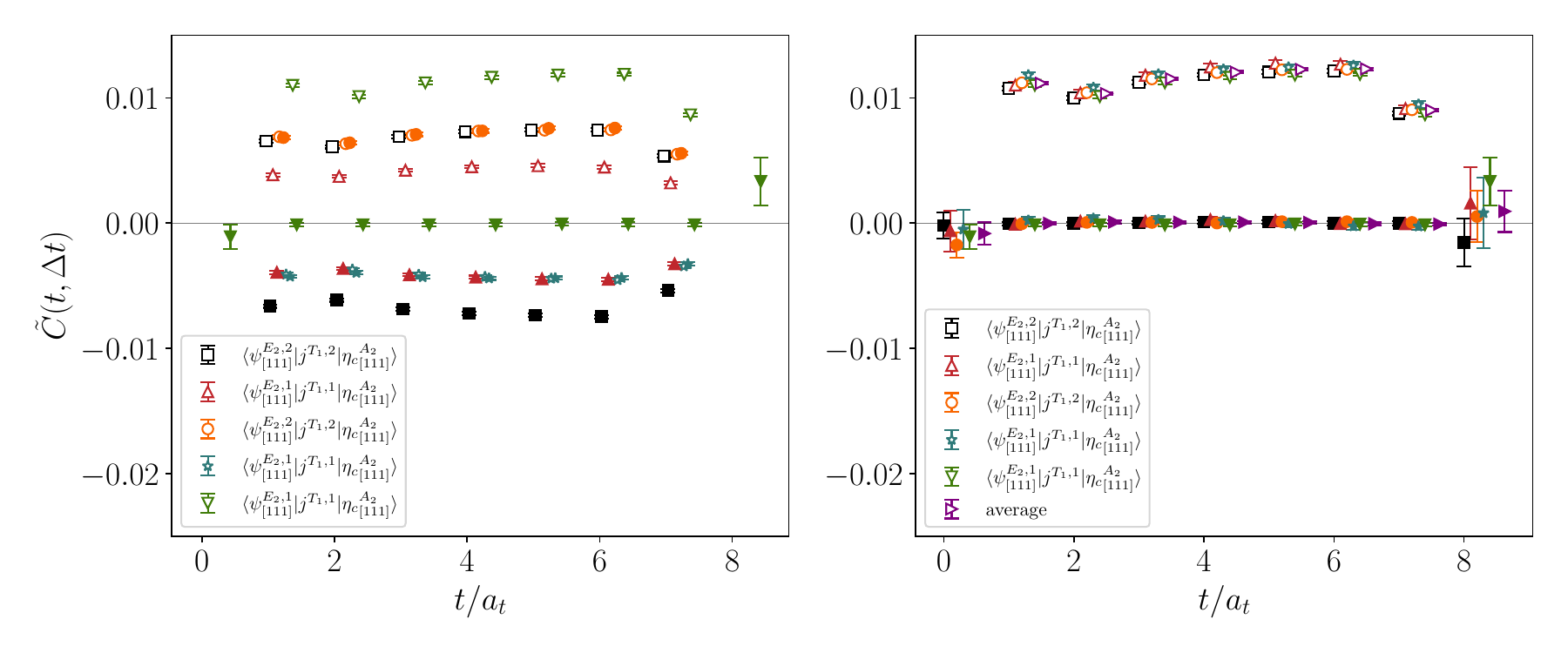}}
  \caption[]{\label{fig:CG-ex-1}
  $\tilde{C}(t, \Delta t = 8 a_t)$ correlation functions for $J/\psi \to \gamma \, \eta_c$ for a single set of irreps. Different symbols reflect the five nonzero row combinations, with open symbols showing the real part and filled symbols the imaginary part. The right panel shows the result of applying the Wigner-Eckart procedure to bring each correlator into a reference configuration (corresponding to the purely real green case in the left panel).  
  }
\end{figure*}

\begin{figure*}
    \centerline{\includegraphics[width=.95\textwidth]{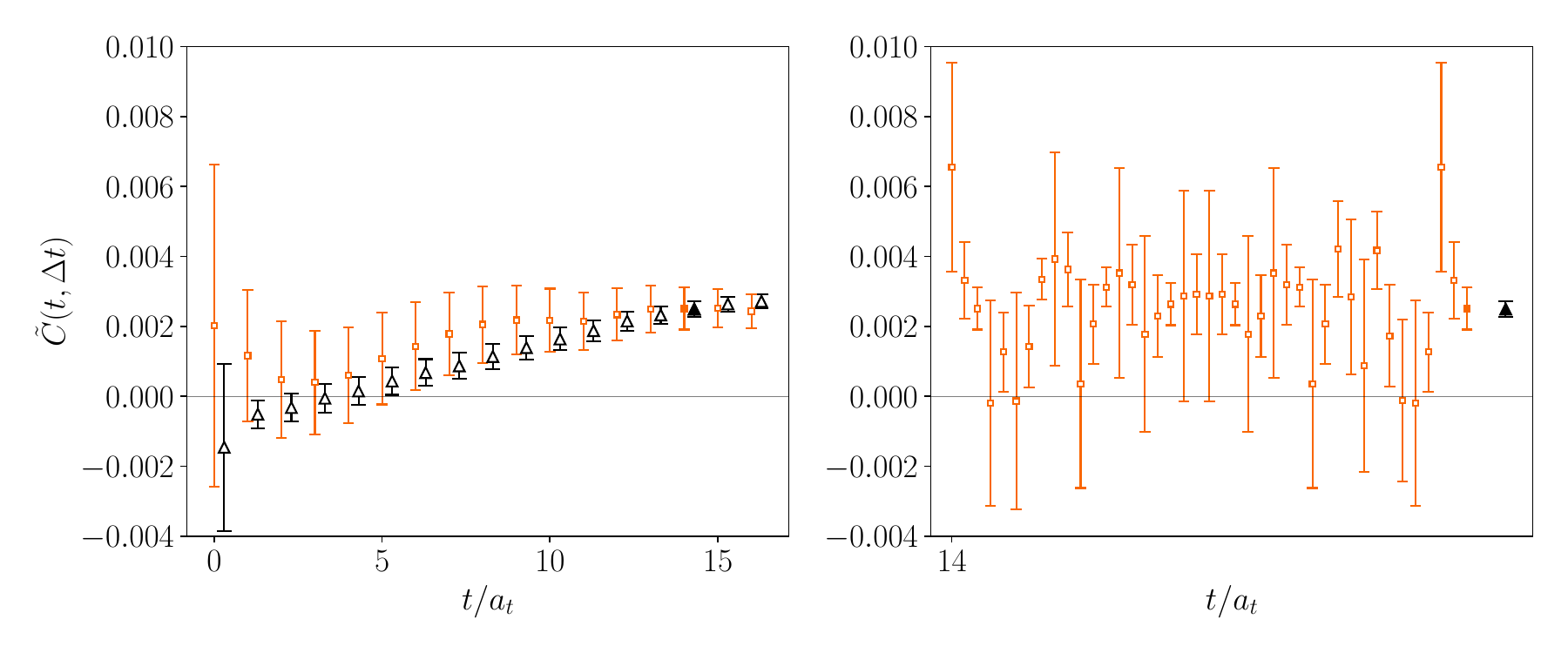}}
    \caption[]{\label{fig:CG-ex-2}
    $\tilde{C}(t, \Delta t = 16 a_t)$ correlation functions for $J/\psi \to \gamma \, \eta$ for a single set of irreps: $\mathbf{n} \Lambda_{\psi} = [\pm 1,\pm 1,\pm 1]\, E_2$, $\mathbf{n} \Lambda_{\eta} = [\pm 1,\pm 1,\pm 1]\, A_2$, and the current at rest in the $T_1^-$ irrep.
      The left panel shows the single reference irrep row combination (orange squares) and the average over all computed correlators after applying the Wigner-Eckart procedure (black triangles).
      The right panel shows the correlator at timeslice $t/a_t=14$ for all irrep row combinations and momentum rotations which go into the averaging procedure after they have been transformed to the reference irrep row, momentum direction combination.
    }
\end{figure*}

To apply the Wigner-Eckart theorem and average the various irrep--row combinations of the three--point functions we need to calculate the appropriate Clebsch-Gordan coefficients.
We will use the method outlined in Ref.~\cite{Dudek:2012gj} with a slight modification.
Consider the Clebsch-Gordan coefficients for $\Lambda' \otimes \Lambda_{\gamma} \to \Lambda$, where $\Lambda', \Lambda_{\gamma}$ and $\Lambda$ are irreps of groups $G',G_{\gamma}$ and $G$ respectively.
The coefficients are calculated using the projection formula,
\begin{align*}
  \label{eq:31}
    \begin{split}
  &\mathcal{C} \big( \{ \mathbf{p} \}^{\star}_{k} \Lambda\mu;\, \{ \mathbf{p}' \}^{\star}_{k'}\Lambda' \mu' ;\, \mathbf{q}\Lambda_{\gamma} \mu_{\gamma} \big) \\
  &= \frac{|\Lambda|}{|G|}  \sum_{R \in G} \Gamma(R)_{k\mu, l\nu}\, \Gamma'(R)^{*}_{k' \mu', l' \nu'}\, \Gamma_{\gamma}(R)_{k_{\gamma}\mu_{\gamma}, k_{\gamma}\nu_{\gamma}} \, ,
\end{split}
\end{align*}
where the $\Gamma$ are the induced representations as defined in Ref.~\cite{Dudek:2012gj}, and $\mu,\, \mu',\, \mu_{\gamma}$ and $\nu,\, \nu',\, \nu_{\gamma}$ label rows of the irreps.
We calculate these coefficients for each set of linearly independent combinations of $\nu,\, \nu',\, \nu_{\gamma}$.
The momenta in the stars are labeled by $k,k'$ and $l,l'$, while $k_{\gamma}$ is defined by $\mathbf{q} = \{\mathbf{q}\}^{\star}_{k_{\gamma}}$; it is sufficient to consider just one particular $l,l'$ pair to obtain all the coefficients.
The rotations $R\in G$ will rotate the momenta such that $\{\mathbf{p}'\}^{\star}_{k'} + \mathbf{q} = \{\mathbf{p}'\}^{\star}_k$ is always satisfied.

\bigskip

To illustrate the effectiveness of averaging over correlation functions that are equivalent under the lattice symmetries, we can consider the radiative transition $J/\psi \to \gamma \, \eta_c$ which has the same Lorentz structure as the radiative decays considered in this paper, but which has much smaller statistical fluctuations owing to it being a connected process \cite{Delaney:2023fsc, Dudek:2006ej}.
Figure~\ref{fig:CG-ex-1} shows $\Delta t/a_t =8$ correlation functions, $\tilde{C}(t, \Delta t)$, computed using optimized $J/\psi$ and $\eta_c$ operators for the case $[1,1,1]\, E_2$, $[1,1,1]\, A_2$ respectively, with the current in the $[0,0,0]\, T_1^-$ irrep. Each possible $J/\psi$ and current row is shown in the left panel, and the result of applying the Wigner-Eckart procedure to transform each into a reference configuration is shown in the right panel. The result confirms that the computed Clebsch-Gordan coefficients correctly represent the lattice symmetry. 
For the example $J/\psi \to \gamma \, \eta_c$ the signal for a single irrep row and momentum direction combination is sufficiently good that the averaging procedure is not required, however when considering the radiative decays in this paper it provides an essential improvement to the signal quality.

Figure~\ref{fig:CG-ex-2} shows the rescaled correlators, ${\tilde{C}(t, \Delta t = 16 a_t)}$ for $J/\psi \to \gamma\,  \eta$ with operators in the same irreps and momenta as the previous example.
The left panel shows the correlator for the (real-valued) reference row--combination (orange squares) and the result of the average over all related momentum directions and irrep rows (black triangles), showing the improvement of the signal achieved with this procedure.
The right panel shows the value of each of the correlators which go into the averaging procedure for this irrep combination, sampled at a single timeslice, $t/a_t = 14$.

\bigskip

\begin{figure}
    \centerline{\includegraphics[width=.99\columnwidth]{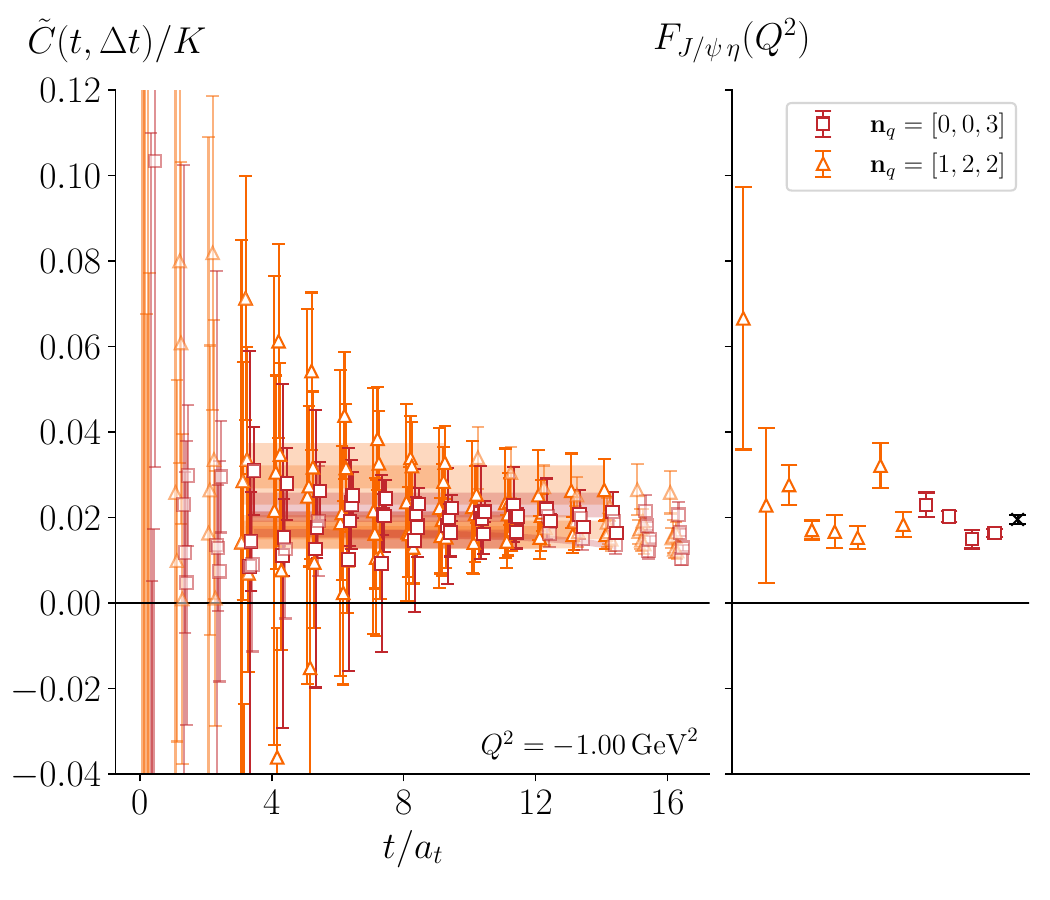}}
  \caption[]{\label{fig:discretization}
    Example of form--factor averaging procedure for $J/\psi \to \gamma \, \eta$.
    The correlators shown with triangles have hadron momenta of $\mathbf{n}_{\psi} = [1,1,0]$ and $\mathbf{n}_{\eta} = [0,\textrm{-}1,\textrm{-}2]$ giving $\mathbf{n}_q = [1,2,2]$.
    The correlators shown with squares have hadron momenta of of $\mathbf{n}_{\psi} = [1,0,1]$ and $\mathbf{n}_{\eta} = [1,0,\textrm{-}2]$ giving $\mathbf{n}_q = [0,0,3]$.
    The black cross on the right shows the averaged value of the form--factor.
    The correlators corresponding to the first two fits in the right panel have been left out of the left panel as they are very noisy (these both have small kinematic factors such that they do not contribute significantly to the final average).
  }
\end{figure}

The second stage of our averaging procedure combines form--factor estimates which would agree if there are negligible discretization effects such that Lorentz symmetry is effectively (approximately) restored.
In Figure~\ref{fig:discretization} we show rescaled correlators at one value of $Q^2$ where the current insertion can have two different values of momenta with the same magnitude, $\mathbf{n}_q=[1,2,2]$ and $\mathbf{n}_q=[0,0,3]$.
Since we expect the form--factor to depend only on the value of $Q^2$, this is a good test for any remaining discretization effects due to the different momenta types and directions.
In the right--hand panel of the figure we show the AIC--averaged value of the fits to each correlator, and while there is a spread in the values, there does not appear to be any systematic difference between the two momentum types of the current insertion.

\section{Analysis Systematics}\label{discretization}
%

\begin{figure}[t]
\centerline{\includegraphics[width=.9\columnwidth]{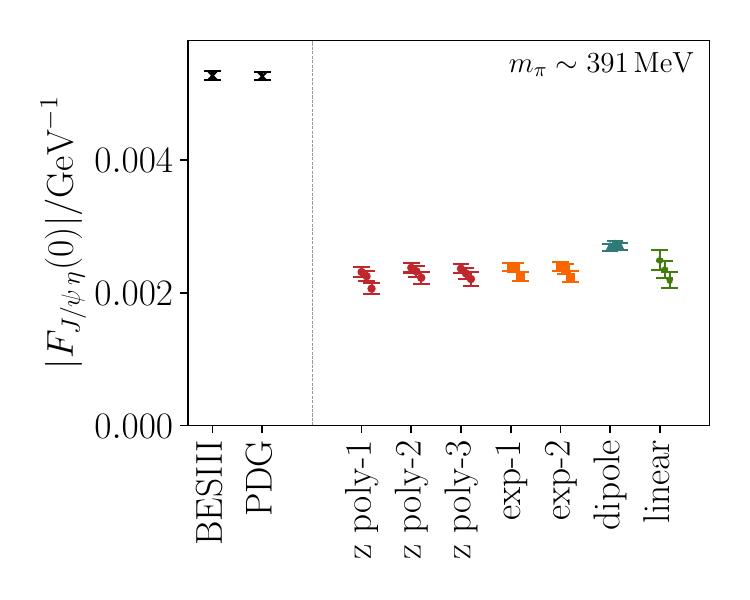}}
\vspace*{-4mm}
\caption[]{\label{fig:exp-sys-eta} Form--factor at $Q^2=0$ for $J/\psi\to\gamma\,\eta$ on this paper's lattice from parameterizations.
The values shown for each fit are in order, the lowest values of $\xi$, the value of $\xi$ used in the paper and the highest value of $\xi$.
}
\end{figure}

\begin{figure}[t]
\centerline{\includegraphics[width=.9\columnwidth]{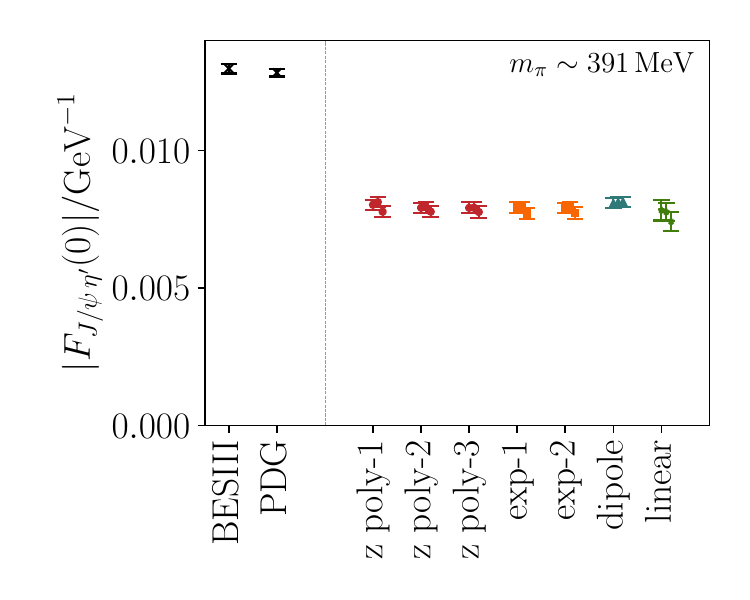}}
\vspace*{-4mm}
\caption[]{\label{fig:exp-sys-eta-prime} As for Fig.~\ref{fig:exp-sys-eta}, but showing the form--factor at $Q^2=0$ for $J/\psi\to\gamma\,\eta'$.
}
\end{figure}

In this appendix we present several sources of systematic uncertainty in the analysis used in this calculation: variation in the value of the anisotropy, $\xi$, choice of source--sink separation, $\Delta t$, and the choice to use only the spatially directed current.

\medskip

The anisotropy of the lattice used in this paper was determined by fitting the relativistic dispersion relation to the spectrum of low-lying charmonium states and $\eta^{(\prime)}$ states, and this resulted in a range of values around $\xi\sim3.5$, we will use the variation in this parameter to determine a systematic uncertainty on the form--factor results, which we will find to be modest.

The value of the anisotropy contributes to the scale setting of $Q^2$--values as well as to the magnitude of the $\mathcal{O}(a)$--improvement to the vector current. We run the full analysis using the lowest ($\xi=3.45$) and highest ($\xi=3.59$) value of the anisotropy, fitting the $Q^2$--dependence and extracting the form--factors for both of these values.

In Figures \ref{fig:exp-sys-eta} and \ref{fig:exp-sys-eta-prime} we show the values of $F(0)$ extracted from the analysis with the lowest value of $\xi$, the value used in the rest of the paper, and the highest value of $\xi$.
We use the variation in the results from the linear fit to points around $Q^2=0$ to determine a systematic uncertainty for the form--factors which we include our best estimates presented in Equation~\ref{FINAL}.

\begin{figure*}
 \begin{subfigure}{\columnwidth}
\centerline{\includegraphics[width=\textwidth]{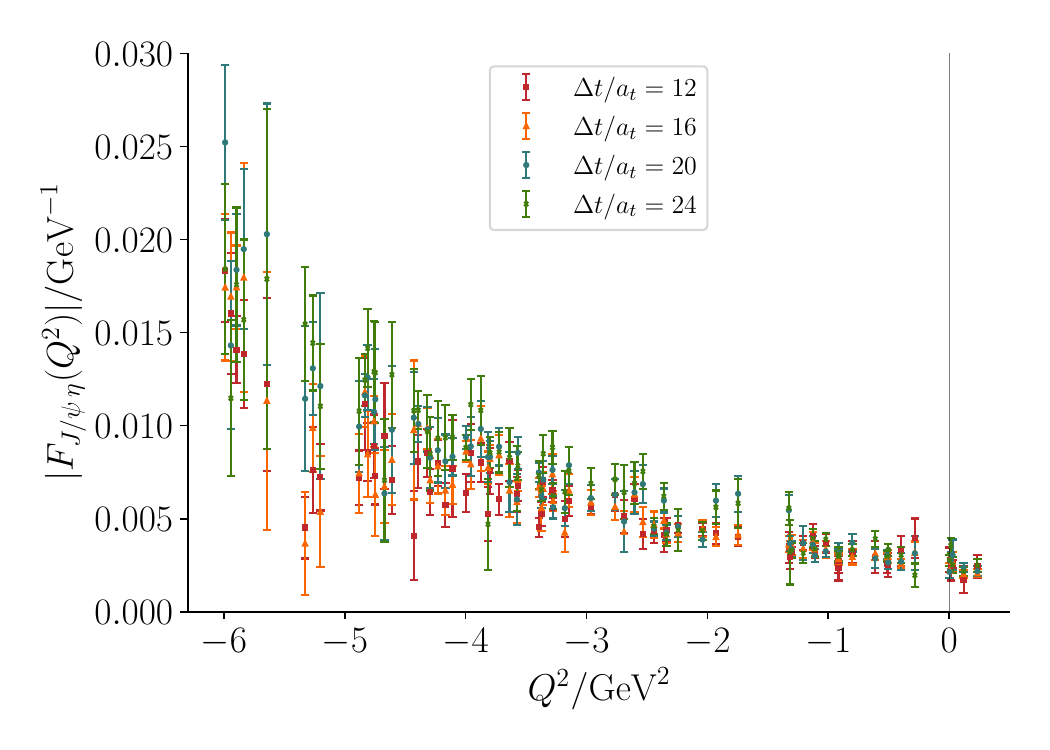}}
\end{subfigure}
  \begin{subfigure}{\columnwidth}
\centerline{\includegraphics[width=\textwidth]{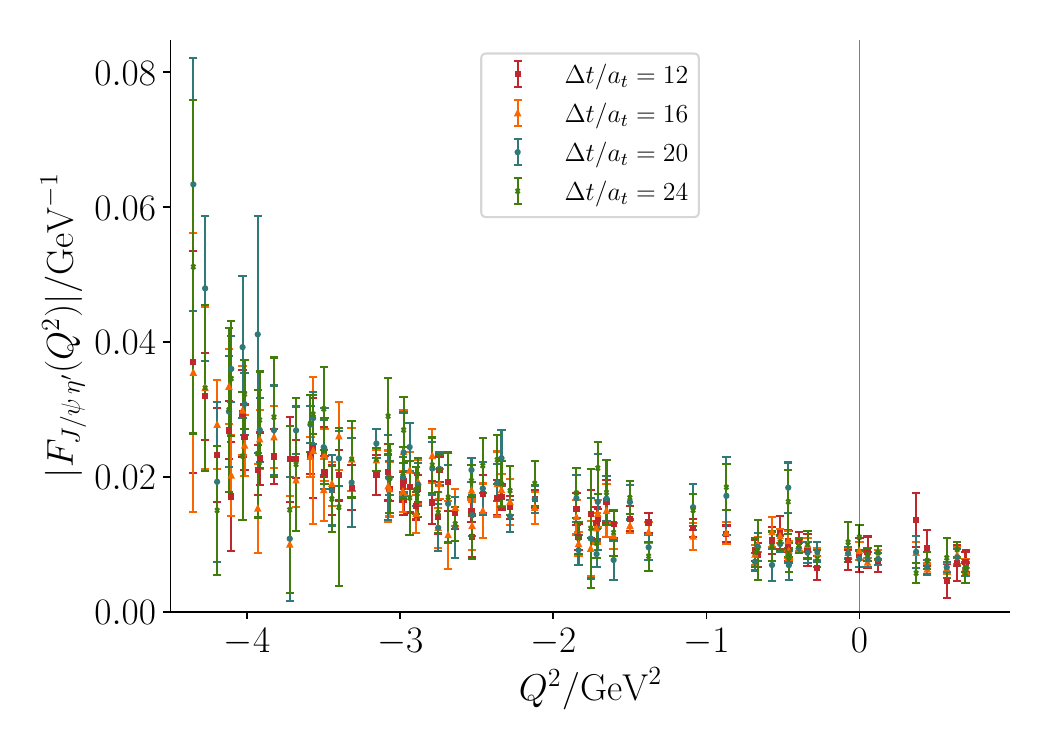}}
\end{subfigure}
\caption[]{\label{fig:ff-dt-together} Form--factors for $J/\psi\to\gamma\, \eta$ (left) and $J/\psi\to\gamma \,\eta'$ (right), showing the results from all four source--sink separations together ($\Delta t/a_t = 12,16,20,24$).}
\end{figure*}

\begin{figure*}
  \begin{subfigure}{\columnwidth}
    \centerline{\includegraphics[width=\textwidth]{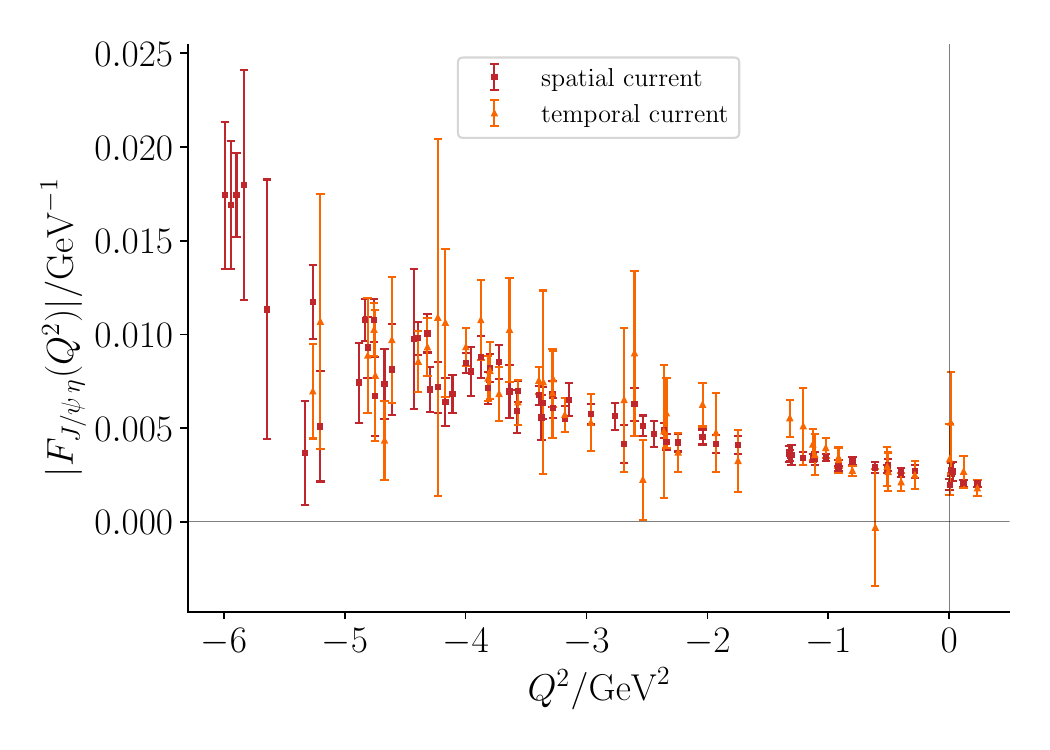}}
  \end{subfigure}
  \begin{subfigure}{\columnwidth}
    \centerline{\includegraphics[width=\textwidth]{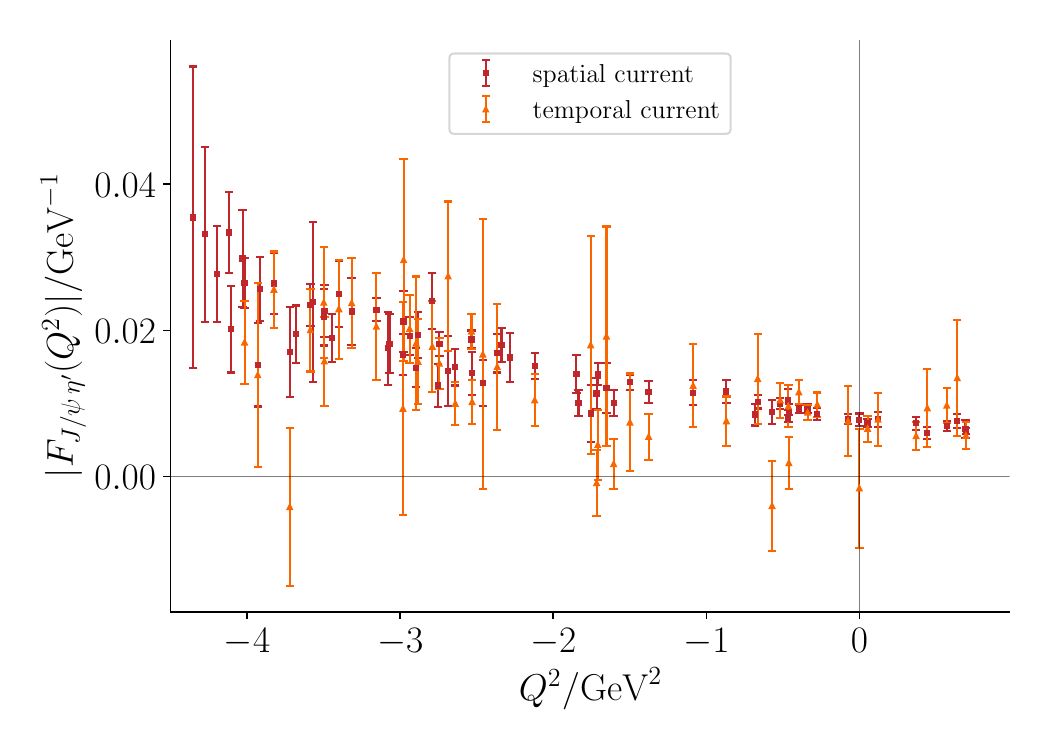}}
  \end{subfigure}
  \caption[]{\label{fig:ff-spat-temp} Form--factor for $J/\psi\to\gamma \, \eta$ (left) and $J/\psi\to\gamma\, \eta'$ (right), showing results for both the improved spatial current and the improved temporal current in the case $\Delta t / a_t = 16$.
  }
\end{figure*}

\bigskip

We computed our full set of correlation functions on choices other than the $\Delta t / a_t = 16$ source--sink separation reported on in the body of the paper. Figure~\ref{fig:ff-dt-together} shows the resulting extractions for $\Delta t/a_t = 12, 16, 20, 24$. We observe that the $\Delta t / a_t = 16$ results are statistically compatible with the less--precise determinations from large values of $\Delta t$. This adds confidence that the AIC--weighted averaging procedure together with the inclusion of additional exponentials in the fit function has allowed us to reliably extract the desired signal with excited state contributions correctly accounted for.

\bigskip

\medskip
The anisotropic Clover--improved lattice used in this paper requires the use of an $\mathcal{O}(a)$-improved vector current, and all results shown in this paper so far have used the improved \emph{spatial} vector current only. To ensure that there is no large systematic difference between the improved spatial and temporal currents we also calculated the form--factors using the temporal current for $\Delta t/a_t = 16$.
In Figure~\ref{fig:ff-spat-temp} we show the form--factor results using the improved spatial and temporal vector currents, which are observed to agree within the large statistical uncertainties on the temporal results. The increased uncertainty is mainly due to the vastly reduced number of irrep and momentum combinations which are non-zero when using the temporal current, removing much of the benefit from averaging over equivalent correlators.


\clearpage\pagebreak
\section{Parameterizations of $Q^2$ Dependence}\label{qsq-fits}

The parameterizations of the $Q^2$ dependence of the form--factors for transitions to the $\eta$ and $\eta'$ discussed in the text are presented here in full detail. In all cases (except for linear fits around $Q^2=0$) the full set of 70 independent data points are included in a correlated $\chi^2$ fit. Data correlations are found to be modest (the data covariance matrix in each case has eigenvalues, $\lambda$, such that $\lambda_\mathrm{min} / \lambda_\mathrm{max} \sim 0.03$).

\noindent\hrulefill

\medskip
\noindent Description with a single dipole form,
\begin{equation*}
\label{eq:11}
F(Q^2) = \frac{F_0}{1+Q^2/\Lambda^2}\, ,
\end{equation*}
with $F_0$ and $\Lambda$ allowed to vary leads to 

\smallskip
\quad{
 \renewcommand{\arraystretch}{1.4}
 \begin{tabular}{rrr}
   & $J/\psi \to \gamma\eta $ & $J/\psi \to \gamma\eta'$\\
   \hline
   $F_0/\textrm{GeV}^{-1}$  & 0.002720(57) & 0.00811(18) \\
   $\Lambda/\textrm{GeV}$  & 2.626(14) & 2.383(25) \\
   \hline
   $\chi^2/N_{\textrm{dof}}$ & 2.1 & 1.1 \\
 \end{tabular}
}
\smallskip 

\noindent as shown by the red curves in Figure~\ref{fig:qsq-fits-dipole-exp}.

\noindent\hrulefill

\bigskip
\noindent Nearby--singularity--free forms based upon exponentials,
\begin{align*}
  F(Q^2) &= F_0\, e^{-Q^2/16\beta^2}, \\
  F(Q^2) &= F_0\, e^{-(1+\alpha Q^2)Q^2/16\beta^2} \, ,
\end{align*}
with $F_0$, $\beta$, and $\alpha$ allowed to vary provide descriptions given by

\smallskip
\quad{
 \renewcommand{\arraystretch}{1.4}
 \begin{tabular}{rrr}
   & $J/\psi \to \gamma\eta $ & $J/\psi \to \gamma\eta'$\\
   \hline
   $F_0/\textrm{GeV}^{-1}$  & 0.002377(65) & 0.00794(19) \\
   $\beta/\textrm{GeV}$  & 0.460(8) & 0.463(10) \\
   \hline
   $\chi^2/N_{\textrm{dof}}$ & 1.00 & 0.81 \\
 \end{tabular}
}

\medskip
\quad{
 \renewcommand{\arraystretch}{1.4}
 \begin{tabular}{rrr}
   & $J/\psi \to \gamma\eta $ & $J/\psi \to \gamma\eta'$\\
   \hline
   $F_0/\textrm{GeV}^{-1}$  & 0.002358(72) & 0.00793(19) \\
   $\beta/\textrm{GeV}$  & 0.450(19) & 0.477(22) \\
   $\alpha/\textrm{GeV}^{-2}$  & 0.01(16) & $-0.020(27)$ \\
   \hline
   $\chi^2/N_{\textrm{dof}}$ & 1.01 & 0.85 \\
 \end{tabular}
}
\smallskip

\noindent as shown by the orange and blue curves in Figure~\ref{fig:qsq-fits-dipole-exp}. 

\noindent\hrulefill

\bigskip
\noindent Mapping $Q^2$ into a variable
\begin{equation*}
z(Q^2) = \frac{\sqrt{t_{\text{cut}} + Q^2} - \sqrt{t_{\text{cut}} - t_{0}}}{\sqrt{t_{\text{cut}} + Q^2} + \sqrt{t_{\text{cut}} - t_{0}}} \, ,
\end{equation*}
with $\sqrt{t_\mathrm{cut}} = 3.043\, \mathrm{GeV}$ and $(\sqrt{t_0})_\eta = 1.926\, \mathrm{GeV}$, $(\sqrt{t_0})_{\eta'} = 1.587\, \mathrm{GeV}$, we considered polynomial descriptions of the data,
\begin{equation*}
F(Q^2) = \sum_{n=0}^k a_n \, z(Q^2)^n \, ,
\end{equation*}
up to cubic order. The resulting fit descriptions are

\smallskip
\quad{
 \renewcommand{\arraystretch}{1.4}
 \begin{tabular}{rrr}
  	& $J/\psi \to \gamma\eta $ & $J/\psi \to \gamma\eta'$\\
   \hline
   $a_0/\textrm{GeV}^{-1}$  & 0.00704(18) & 0.01627(45) \\
   $a_1/\textrm{GeV}^{-1}$  & $-0.0370(15)$ & $-0.1029(56)$ \\
   \hline
   $\chi^2/N_{\textrm{dof}}$ & 1.07 & 1.02 \\
   $|F(0)|/\textrm{GeV}^{-1}$ & 0.002247(75) & 0.00812(19)\\
 \end{tabular}
}
\smallskip

\smallskip
\quad{
 \renewcommand{\arraystretch}{1.4}
 \begin{tabular}{rrr}
   & $J/\psi \to \gamma\eta $ & $J/\psi \to \gamma\eta'$\\
   \hline
   $a_0/\textrm{GeV}^{-1}$  & 0.00716(18) & 0.01638(45) \\
   $a_1/\textrm{GeV}^{-1}$  & $-0.0447(32)$ & $-0.1342(99)$ \\
   $a_2/\textrm{GeV}^{-1}$ & 0.056(21) & 0.348(90) \\
\hline
   $\chi^2/N_{\textrm{dof}}$ & 0.98 & 0.81 \\
      $|F(0)|/\textrm{GeV}^{-1}$ & 0.002323(80) & 0.00793(19)\\
 \end{tabular}
}
\smallskip

\smallskip
\quad{
 \renewcommand{\arraystretch}{1.4}
 \begin{tabular}{rrr}
   & $J/\psi \to \gamma\eta $ & $J/\psi \to \gamma\eta'$\\
   \hline
   $a_0/\textrm{GeV}^{-1}$  & 0.00705(20) & 0.01651(59) \\
   $a_1/\textrm{GeV}^{-1}$  & $-0.0451(32)$ & $-0.135(10)$ \\
   $a_2/\textrm{GeV}^{-1}$ & 0.103(45) & 0.28(22) \\
   $a_3/\textrm{GeV}^{-1}$ & $-0.30(26)$ & 0.6(19) \\
\hline
   $\chi^2/N_{\textrm{dof}}$ & 0.97 & 0.82 \\
      $|F(0)|/\textrm{GeV}^{-1}$ & 0.002290(84) & 0.00792(20)\\
 \end{tabular}
}
\smallskip

\noindent as shown by the orange, blue and green curves in Figure~\ref{fig:qsq-fits-zexp}.

\noindent\hrulefill

\bigskip

\noindent Description of a limited region around $Q^2=0$ as shown in the insets of Figure~\ref{fig:ff-snk-highlight} is achieved with a simple linear interpolation,
\begin{equation*}
F(Q^2) = F_0 + s\,  Q^2 \, ,
\end{equation*}
with fit parameters,

\smallskip
\quad{
 \renewcommand{\arraystretch}{1.4}
 \begin{tabular}{rrr}
   & $J/\psi \to \gamma\eta $ & $J/\psi \to \gamma\eta'$\\
   \hline
   $F_0/\textrm{GeV}^{-1}$  & 0.00235(13) & 0.00777(33) \\
   $s/\textrm{GeV}^{-3}$  & $-0.00156(96)$ & $-0.0020(16)$ \\
\hline
   $\chi^2/N_{\textrm{dof}}$ & 1.24 & 0.18 \\
 \end{tabular}
}


\end{document}